\documentclass[aps,pre,groupedaddress,twocolumn]{revtex4-2}

\usepackage{amsmath}    
\usepackage{graphicx}   
\usepackage{verbatim}   
\usepackage{color}      
\usepackage{subfigure}  
\usepackage{hyperref}   
\newcommand{\bq}{\begin{equation}}
\newcommand{\eq}{\end{equation}}
\newcommand{\bqn}{\begin{eqnarray}}
\newcommand{\eqn}{\end{eqnarray}}
\newcommand{\nb}{\nonumber}
\newcommand{\lb}{\label}

\newcommand{\jcap}{JCAP}
\newcommand{\mnras}{MNRAS}
\newcommand{\aj}{AJ}
\newcommand{\physrep}{Phys. Rep.}

\begin{document}


\title{Assessment of Dark Matter Models Using Dark Matter Correlations across Dwarf Spheroidal Galaxies}

\author{Ahmad Borzou}
\email[]{ahmadborzou@gmail.com}
\affiliation{EUCOS-CASPER, Department of Physics, Baylor University, Waco, TX 76798, USA}


\date{\today}

\begin{abstract}
The predicted size of dark matter substructures in kilo-parsec scales is model-dependent. Therefore, if the correlations between dark matter mass densities as a function of the distances between them are measured via observations, we can scrutinize dark matter scenarios. In this paper, we present an assessment procedure of dark matter scenarios. First, we use Gaia's data to infer the single-body phase-space density of the stars in the Fornax dwarf spheroidal galaxy. The latter, together with the Jeans equation, after eliminating the gravitational potential using the Poisson equation, reveals the mass density of dark matter as a function of its position in the galaxy. We derive the correlations between dark matter mass densities as a function of distances between them. No statistically significant correlation is observed. Second,
for the sake of comparison with the standard cold dark matter, we also compute the correlations between dark matter mass densities in a small halo of the Eagle hydrodynamics simulation. We show that the correlations from the simulation and from Gaia are in agreement. Third, we show that Gaia observations can be used to limit the parameter space of the Ginzburg--Landau statistical field theory of dark matter mass densities and subsequently shrink the parameter space of any dark matter model. As two examples, we show how to leave limitations on (i) a classic gas dark matter and (ii) a superfluid dark matter.
\end{abstract}


\maketitle

\section{Introduction}
\lb{Sec:intro}

At the moment, the widely used cosmological model consists of a constant $\Lambda$ accounting for the vacuum energy, a cold collisionless dark matter (CDM), visible matter, and general relativity. Despite the popularity of $\Lambda$CDM, there exist tensions between its predictions and observations. The discrepancy in the value of the Hubble constant 
 \cite{2021CQGra..38o3001D}, the impossibly early galaxy problem \cite{2016ApJ...824...21S,2019EPJC...79..571Y}, and the high-z quasar problem \cite{2018EL....12339001M} can be mentioned as the challenges of the $\Lambda$CDM model. 

In the dark matter (DM) sector of $\Lambda$CDM, at scales larger than 1 mega-parsecs, the observations are consistent with CDM. The cosmological model based on CDM provides a fairly accurate description of galaxy evolution, galaxy counts, and even galaxy morphology~\cite{2014MNRAS.444.1518V, 2015MNRAS.446..521S}. Nevertheless, there are observations at the galactic scales that are hard to understand in the context of CDM. 
The observed mass densities of DM at the center of galaxies are (i) shallower \cite{Alam:2001sn, 2015MNRAS.452.3650O} and (ii) less steep \cite{2005ApJ...621..757S,2001AJ....122.2381M} than predicted by CDM cosmology. Therefore, CDM predictions of (i) the mass density of DM and (ii) the first derivative of the mass density are in disagreement with observations. 
Another class of observations that seem to contradict the predictions of CDM is related to the number of observed subhalos in galaxies such as our own Milky Way. While we have observed only $\sim$50 satellite or dwarf galaxies within the Milky Way, CDM predicts the number to be around 1000 \cite{2008MNRAS.391.1685S}.  
Although some of these faint objects may have not been discovered, the difference between the observed and predicted counts is significantly high. 
Moreover, many of the observed satellite galaxies have total halo masses much less than the heaviest subhalos predicted by CDM. It is hard to understand why the heavier subhalos have failed to form galaxies while the less massive subhalos with lower efficiency in star formation have been observed \cite{2014MNRAS.440.3511T}. 

Two classes of alternative scenarios have been introduced to solve the above mentioned small-scale problems. In the first category, baryonic feedback within the CDM framework accounts for the discrepancies \cite{2008Sci...319..174M}. In the second category, modified models of DM are suggested. Warm DM, self-interacting DM, 
degenerate fermionic DM, Bose--Einstein condensed models of DM, and superfluid DM are among the scenarios that are proposed for solving the small-scale problems \cite{doi:10.1146/annurev-astro-082708-101659}.

Which one of these scenarios is better? By design, all of the proposals have a chance of explaining the mentioned small-scale problems while behaving more or less like CDM on larger scales. We need additional experiments or new analyses of the data from the existing experiments that can assess the DM scenarios in the domains that are independent of the mean density of dark matter. The present article is a contribution to the latter direction.

In this paper, we introduce a procedure to estimate the correlations between DM mass densities across dwarf spheroidal (dSph) galaxies and use that to lay limitations on DM models. We first learn the single-body phase-space density of the stars in the dSph galaxies from the
motion of their stars. Next, we take a divergence of the Jeans equation and combine it with the Poisson equation to write the mass density distribution of DM in terms of the estimated single-body phase-space density of the stars. Estimation of DM mass density using the single-body phase-space density of stars has been reported in several publications. For example, see \cite{2000MNRAS-313-209H, 2016MNRAS-459-4191S} and the references therein. We use the estimated mass densities to explore the correlations between them. The estimated correlations shall be explained by any proposed DM scenario and results in a reduction of its parameter space. Since the correlations are independent of the mean DM mass distributions, as we show in Appendix~\ref{App:InteractiveGas}, the imposed restrictions are in addition to the requirement of explaining the small-scale observations.

There are two ways to use the measured correlations to leave limitations on a DM model. (i) In the case of sophisticated models, high resolution N-body simulations can be used to find the predicted correlations between mass densities across a small halo. The predicted correlations can then be compared with the measured correlations from observations. In this paper, we use the DM hydrodynamics simulations in the Eagle project to show that their CDM simulations produce no correlation between DM mass densities across small halos. (ii) In the case of simple analytic DM models, assuming that the halo exchanges DM with the surroundings to prevent the gravothermal catastrophe, we use the Ginzburg--Landau approach to construct the statistical field theory of the mass densities by expanding the free energy of the halo, in its general form, around the observed mean mass densities. The steadiness of the halo guarantees the smallness of the higher order terms. By neglecting these terms, we are able to straightforwardly derive an expression for the mass density correlations in terms of the coefficients of the free energy expansion. By comparing the correlations that are inferred from observations with the predicted correlations, one can place limitations on the coefficients of the free energy expansion. Since these coefficients are related to the underlying physics of a given model of DM, one can place bounds on the parameter space of the model.

As a showcase, we apply our procedure to the observations of the Fornax dSph galaxy, collected by the Gaia experiment. We observe no statistically significant correlation between DM mass densities that are apart by at least 100~(pc). We use this result to shrink the parameter space of (i) a classic DM gas and (ii) a superfluid DM as two examples of proposed DM models. It should be emphasized that the validity of these results depend on a few assumptions that are made to compensate the lack of observations of the z-components of positions and velocities of stars in dSph galaxies in the Gaia dataset. Therefore, this paper is more a presentation of a DM model assessing procedure than a full analysis of data. When the Gaia limitations are elevated, by, for example, integration of the results of other experiments, a reliable analysis will be possible. This full analysis is left for future works.

This paper is structured as follows. In Section~\ref{Sec:RealizeDM_From_f}, we review the theoretical framework for deriving the correlations between mass densities of DM from the motion of stars in dSph galaxies. In the same section, we derive the theoretical form of the mass density correlations starting from the free energy of the model. In Section~\ref{Sec:Eagle}, we estimate the DM correlations in a simulated small halo of the Eagle project. In Section~\ref{Sec:Data}, we retrieve the observations of the stars in Fornax dSph from Gaia and feed them into the theoretical framework and present the results. In the same section, a few DM models are assessed. A conclusion is drawn in Section~\ref{Sec:Conclusion}.

\section{Theoretical Layout}
\lb{Sec:RealizeDM_From_f}
We begin with the widely used assumption that a given dSph galaxy has reached a steady state; see, for example, \cite{2015JCAP...01..002D,  2016MNRAS.461.2914H,  2018RPPh...81e6901S,  2018MNRAS.475.5385D}. In the case of the Fornax dSph, this assumption will be validated by data later in this article. Therefore, we start from the Jeans equation for the stars in the galaxy
\bqn
\lb{Eq:JeansEq}
\frac{1}{\rho^*}\partial_j \left(\rho^* \sigma^{2* i j} \right)+ \partial^i \phi = 0,  
\eqn
where an asterisk refers to the visible matter in the galaxy, $\phi$ is the gravitational potential, and the mass density and the dispersion velocity of the visible matter respectively read
\bqn
\lb{Eq:rho_sig_vismatter}
&&\rho^* = m^* \int d^3v f^* , \nb\\
&&\sigma^{2* i j} = \int d^3v f^* v^i v^j, 
\eqn
where $v$ is the velocity of stars, $f^{*}$ is the one-body phase-space density of stars, and $m^*$ is the granular mass of the stars, which will be canceled out later in the calculations.

After taking a divergence of Equation~\eqref{Eq:JeansEq}, and using the Poisson equation, the mass density of DM in the galactic halo reads
\bqn
\lb{Eq:DMMassDensityInTermsOfVisible}
\rho = -\rho^* - \frac{1}{4\pi G}\partial_i\left(\frac{1}{\rho^*} \partial_j \left(\rho^* \sigma^{* i j} \right) \right),
\eqn
where $G$ is the gravitational constant, and the mass $m^*$ will be canceled out in the second term. As far as the dwarf spheroidal galaxies are concerned, we can neglect the first $\rho^*$ on the right-hand side of this equation, and the DM mass density is approximately equal to the second term. Therefore, if we estimate the one-body distribution function $f^*$ from observations, the mass density of DM is known in terms of the positions in the galaxy. 

\subsection{Density Correlations from Observations}
Assuming that DM mass density $\rho$ has been estimated, we first define the mean of DM mass density as 
\bqn
\lb{Eq:rho_bar}
\bar{\rho} \equiv \frac{1}{\int d^3x} \int d^3x\, \rho(x),
\eqn
where the integration is across a relatively large volume around the center of the halo. We define the DM mass density fluctuations as 
\bqn
\lb{Eq:varphi_Def}
\varphi(x) \equiv \frac{\rho(x) - \bar{\rho}}{\bar{\rho}}. 
\eqn

The correlations 
 between the density fluctuations of DM, separated by distance $\Delta$, can be estimated from observations 
\bqn
\lb{Eq:FlucCorr}
C(\Delta) = \frac{1}{\int d^3x\oint d\Omega_{\Delta}} \int d^3x\oint d\Omega_{\Delta}\, \varphi(\vec{x})\varphi(\vec{x}+\vec{\Delta}), 
\eqn
where $\oint d\Omega_{\Delta}$ means integration over all the directions of $\vec{\Delta}$.

\subsection{Density Correlations from DM models}
\lb{Subsec:TheoryDensityCorrs}
We assume that the DM halo has reached a steady state after a long period of evolution. Moreover, we assume that the halo can exchange dark particles with its host and with the cosmic DM background and consequently obeys the statistics of a grand canonical ensemble with a partition function equal to ${\cal{Z}} = \sum_{E,  N} \exp\left(-\beta \left(E-\mu N\right) \right)$, where $E$ and $N$ are the energy and the number of particles of the halo, $\beta$ is the inverse of the temperature, and $\mu$ is the chemical potential due to the exchange of dark particles with the surrounding. The presence of $\mu$ in this statistics guarantees a state of minimum free energy and prevents the gravothermal catastrophe, which occurs in gravity dominant systems with a conserved number of particles. See Appendix~\ref{App:InteractiveGas} for more details. The partition function can be rearranged into the following form; see one such rearrangement for a trivial case in Appendix~\ref{App:CollisionLess_FreeEnergy}, 
\bqn
\lb{Eq:GeneralPartitionFunction}
{\cal{Z}} = \int D\rho\,e^{-\beta F[\rho]},
\eqn
where $F[\rho]$ is the free energy functional, and $\int D\rho$ is the path integral over all possible DM density configurations.

Since the halo is in the steady state, i.e., $\frac{\delta F}{\delta \rho}|_{\bar{\rho}}=0$, the free-energy functional can be expanded around the mean field $\bar{\rho}$ to write the partition function as (see \cite{1991PhRvA..43.2870O, 2015PhR...572....1H} for practical examples)
\bqn
\lb{Eq:GeneralPartionFunc}
{\cal{Z}} =
\int D\varphi\, \exp\left(-\frac{\beta}{2} \int d^3x  \left( \varphi \,\frac{\delta^2 F}{\delta \rho^2}|_{\bar{\rho}} \,\varphi + {\cal{O}}(\varphi^3) + \varphi J   \right)\right),\nb\\
\eqn
where $\frac{\delta^2 F}{\delta \rho^2}|_{\bar{\rho}} =  \gamma \partial^2 + \nu^2$ is the inverse of the Greens' function ${\cal{G}}^{-1}$ and $\gamma$ and $\nu^2$ are free parameters to be determined by the underlying physics. We have added an extra term $\varphi J$ to be set to zero later on and have set $D\rho = D\varphi$ after ignoring a normalization factor. In this paper, due to low data statistics, we ignore the higher order terms. In Appendix~\ref{App:InteractiveGas}, starting from the microscopic model of simple gases, the corresponding $\gamma$ and $\nu^2$ of that model are derived. Derivation of these coefficients in terms of the physics of other DM scenarios is left for the future. Since the higher order terms are ignored, the partition function can be calculated analytically and reads
\bqn
\lb{Eq:Analytic_PartitionFunction}
{\cal{Z}} =\exp\left(\frac{\beta}{2} \int d^3x \int d^3y J(\vec{x}) {\cal{G}}(\vec{x}-\vec{y}) J(\vec{y}) \right),
\eqn
where the normalization factor is dropped. To arrive at this equation, one needs to perform a linear transformation that diagonalizes the inverse Greens' function in Equation~\eqref{Eq:GeneralPartionFunc}. Next, the path integration can be separated into multiplications of independent one-dimensional Gaussian integrals with known answers.

The correlation between the DM fluctuations reads; see, for example, \cite{DavidTongStatFieldTheory}
\bqn
\lb{Eq:Theoretical_Correlation}
C(\Delta) &=& \beta^{-2}{\cal{Z}}^{-1} \frac{\delta^2 {\cal{Z}}}{\delta J(\vec{x}) \delta J(\vec{x}+\vec{\Delta})}\Big|_{J=0}
=\beta^{-1} {\cal{G}}(\Delta)
\nb\\
&=& 4\pi \left(\beta \gamma \Delta\right)^{-1}\exp\left(-\Delta /\sqrt{\beta\gamma/\beta\nu^2}\right).
\eqn
This correlation function should be compared with the one estimated from observations in Equation~\eqref{Eq:FlucCorr}. To arrive at this equation, we note that the correlation is the weighted mean of the multiplication of the fluctuations $\langle \varphi(\vec{x})\varphi(\vec{x}+\vec{\Delta})\rangle \equiv {\cal{Z}}^{-1} \int D\varphi\, \varphi(\vec{x})\varphi(\vec{x}+\vec{\Delta}) \, e^{-\beta F}$. This expression of the correlation function is achieved through the second term in Equation~\eqref{Eq:Theoretical_Correlation} with the partition function given by Equation~\eqref{Eq:GeneralPartionFunc}. If instead, we use the partition function in Equation~\eqref{Eq:Analytic_PartitionFunction} and take the two functional derivatives, the third term of Equation~\eqref{Eq:Theoretical_Correlation} would be the result. Since, by definition, $(\gamma \partial^2 + \nu^2){\cal{G}}(\Delta) = \delta^3(\vec{\Delta})$, the last term of Equation~\eqref{Eq:Theoretical_Correlation} has the form of a Yukawa potential.

It should be mentioned that, as far as the halo is in the steady state, the form of the partition function in Equation~\eqref{Eq:GeneralPartionFunc} is independent of the details of the DM model and the coefficients $\gamma$,  $\nu^2$, and those of the higher order terms are the fingerprints of the model. In principle, if enough high precision data are collected, we would be able to estimate the coefficients up to sufficiently high orders and construct the true model of DM from data with no further assumption. The recipe would be to use the data to estimate the correlation function of Equation~\eqref{Eq:FlucCorr} as well as higher order correlation functions and then solve a system of equations to derive the coefficients of the free energy expansion. Unfortunately, the precision and low statistics of current experiments do not allow such a data-driven model building approach.

\section{Showcase I: Eagle Simulation}
\lb{Sec:Eagle}

In this section, we treat the CDM simulation of the Eagle project \cite{2015MNRAS.446..521S,  2015MNRAS.450.1937C} as if it is the real data. Our goals are to (i) extract the DM correlations predicted by CDM such that we can test them later when they are understood via actual data and (ii) present the potentials of the theoretical method of Section~\ref{Sec:RealizeDM_From_f}.   
  
There are multiple simplifications when working with simulations. First, in actual data, we inevitably extract the mass density of DM from visible matter component. In simulation, the DM information is directly given. Second, an experiment like Gaia does not provide the z-components of the stars in dSph galaxies while these are known for all of the particles in simulations. Third, the systematic errors of star observations are large in an experiment like Gaia. When propagated to the DM sector, they become even larger. The systematic errors are absent in the simulations and we only need to deal with the statistical errors, which, as we see below, are quite small.

We use the particle data of the Eagle simulation with reference name RecalL0025N0752 at zero redshift and select a relatively small halo with GroupNumber = 1 and SubGroupNumber = 123  and halo mass of $\sim 10^9$~M$_\odot$.  
We retrieve the DM particle's positions in the halo using the Python code snippets provided in \cite{2017arXiv170609899T}. In the following, we use the positions of DM particles to estimate the DM mass density $\rho$ and subsequently estimate the correlations. The whole process as well as statistical error estimation is implemented in a Python code that is publicly available at \cite{GaiaPythonCodeBorzou}. The code is intended to work with Gaia dataset, but the following procedures can be achieved with a minimal change.

We convert $(x, y, z)$ to dimensionless variables by dividing each of the components by the standard deviation of the corresponding component of all the DM particles in the halo. Next, we use the kernel density estimator \cite{10.1214/aoms/1177728190,2012arXiv1212.2812Z} to estimate $\rho$ in the three dimensional position-space. In this method, every particle contributes a Gaussian weight to a given point in the position-space. The sum of all the particles' contributions to that point will be the probability of DM mass density there. In other words, the single-body phase-space density reads
\bqn
\rho(\vec{x}) = M {\cal{N}}\sum_{i} \exp\left(-\frac{|\vec{x}-\vec{x}_{i}|^2}{h^2}\right),
\eqn
where $M$ is the total mass of the halo known through the simulation, ${\cal{N}}$ is the normalization factor and is set such that the position integral of the density is equal to $M$, $i$ enumerates the DM particles in the halo, and $h$ is a free parameter to be determined such that the error is minimal. We use the implementation of the method in scikit-learn library, the neighbors class, and the KernelDensity method of the Python programming language. Ideally, we would like to explore the smallest lengths in a given halo, which requires small $h$. Nevertheless, for a fixed number of particles in the halo, there is an optimal $h$ that minimizes the error in $\rho$ estimation but is greater than the ideal value. In general, better resolutions require higher number of particles in the simulation. To estimate the optimal $h$, we use the GridSearchCV method of model\_selection class of the scikit-learn library to explore the parameter space. See \cite{Heidenreich2013} for a description of the method. We find that the optimal $h$ is equal to $0.9$, which, when scaled back to the position-space, is equivalent to $\sim$3~(kpc).

To estimate the statistical error, we use the estimated $\rho$, which serves as the probability of finding DM particles at a given position, to draw a random sample dataset of the same size as the original one. Next, we estimate $\rho$ from the generated dataset by repeating the whole process above. We repeat the sampling and $\rho$ estimation until the standard deviation of the estimations of $\rho$ reaches steady values. The stable standard deviations are assigned as the statistical errors.

Finally, we use Equations~(\ref{Eq:rho_bar}) and (\ref{Eq:varphi_Def}) to compute the fluctuations $\varphi(\vec{x})$, and substitute them into Equation~\eqref{Eq:FlucCorr} to estimate the correlations between fluctuations as a function of the distance $\Delta$. The two-point correlation predicted by CDM can be seen in Figure~\ref{Fig:Correlation_Eagle}. As the figure indicates, the statistical errors are relatively small and no significant correlation is predicted. Later, in Section~\ref{Sec:Data}, we analytically derive this result for a cold classic gas of DM, which is the underlying assumption of CDM simulations.

\begin{figure}
\hspace{-4mm}\includegraphics[width=\columnwidth]{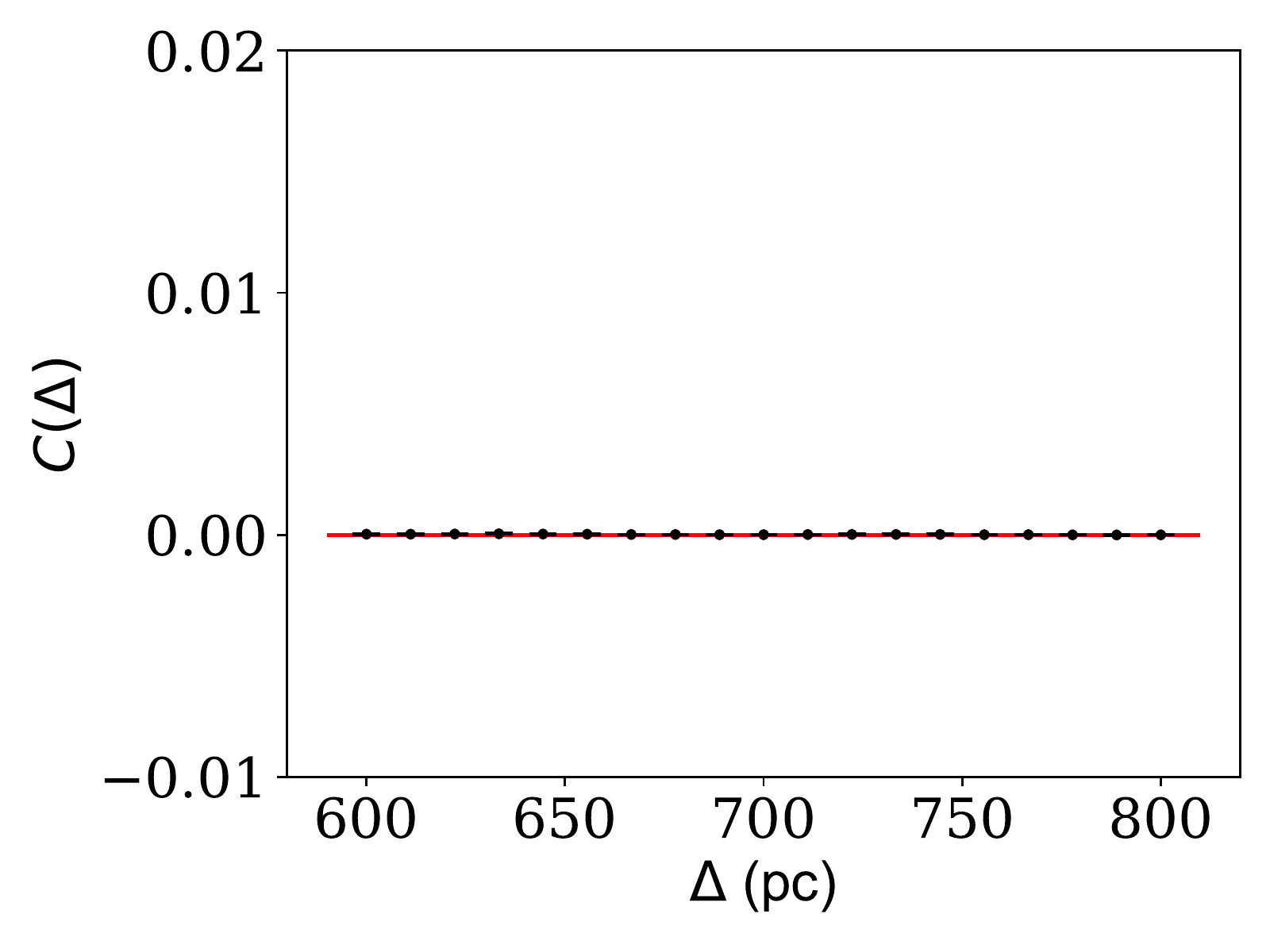}
\caption{We use the hydrodynamic 
 simulations of the Eagle project and select a halo of mass $10^9$~M$_{\odot}$ at zero redshift. We estimate the DM mass density across the halo and use that to find the two-point correlations $C(\Delta)$ between DM mass density fluctuations as a function of the distance between the position of mass densities $\Delta$. We observe no statistically significant correlation. The statistical errors are estimated and shown in the figure. \lb{Fig:Correlation_Eagle}}
\end{figure}

\section{Showcase II: Gaia Data}
\lb{Sec:Data}
In this section, we use the observations made by Gaia to learn the positions and velocities of the stars belonging to the Fornax dSph galaxy. After filtering and processing the data, we insert the results into the theoretical framework of Section~\ref{Sec:RealizeDM_From_f} to derive the mass density distribution of DM and the correlations between them.  
The whole process is implemented in a Python code that is publicly available \cite{GaiaPythonCodeBorzou}.  

It should be emphasized that, in Section~\ref{Subsec:DensityEstimation}, we will compensate for the limitations of the Gaia experiment with an assumption whose validity is not known with certainty. Therefore, the results are only valid to the zeroth-order approximation for $f^*$. In the future, one can integrate other observations with the Gaia data to remove the limitations. In that case, the assumptions made in this section are unnecessary and a thorough analysis will be possible.

\subsection{Selecting Stars}
Billions of stars have been observed by Gaia. Among them, we need to extract those stars that belong to the Fornax dSph. For the Fornax dSph member selection, we use GetGaia, a Python code published in \cite{2021ApJ...908..244D, 2021MNRAS.505.5884M}, which performs a series of screening tasks on the dataset made available by Gaia. GetGaia starts with the stars within a large enough spherical region around the galaxy. Those stars with poor color and astrometric solutions are dropped. Next, those stars whose proper motion or parallax is not consistent with the bulk of the galaxy are filtered out. The selected stars have errors of less than 0.5 mas in parallax, less than 0.5 mas per year in proper motion, less than 0.1 mag in both integrated BP mean magnitude and integrated RP mean magnitude, and less than 0.01 mag in the G-band mean magnitude. Moreover, the stars with integrated BP/RP mean magnitudes of higher than 20.5 are removed. Since star selection and contamination removal in our work is entirely done by GetGaia, we refer to \cite{2021ApJ...908..244D, 2021MNRAS.505.5884M} for a more detailed description of the selections above and the reasons behind them. 

\subsection{Coordinate System}
\lb{Subsec:Coordinate}
The calculations of Section~\ref{Sec:RealizeDM_From_f} can be best carried out in a comoving Cartesian coordinate system $(x, y, z)$ whose origin is at the center of the Fornax dSph. The position and velocity of a star in this frame are
\bqn
\vec{x} = \vec{r} - \vec{r}_m,\nb\\
\vec{\dot{x}} = \vec{v}-\vec{v}_m,
\eqn
where $\vec{r}$ refers to the position of the star in a spherical coordinate system $(r, \theta,\phi)$ attached to the sun with the polar and azimuthal angles defined as $\theta = \frac{\pi}{2} - \text{dec}$ and $\phi = \text{ra}$, where dec and ra stand for the declination and right ascension of the equatorial coordinate system provided by Gaia. The subscript $m$ refers to the center of the galaxy in the spherical system, and $\vec{r}_m$ is defined as the mean of the position of the stars. Initially, the unit vectors of the co-moving coordinate system are set equal to the unit vectors of the spherical system at angles $\theta_m$ and $\phi_m$ that define $\vec{r}_m$
\bqn
&&\hat{x}\equiv\hat{\theta}(\theta_m,\phi_m), \nb\\
&&\hat{y}\equiv\hat{\phi}(\theta_m,\phi_m),  \nb\\
&&\hat{z}\equiv\hat{r}(\theta_m,\phi_m).
\eqn 

The components of the position and velocity of a star at $(r, \theta, \phi)$, in the comoving coordinate system reads
\bqn
&&x = \vec{r} \cdot \hat{x} \simeq  D\, \hat{r}(\theta,\phi)\cdot \hat{\theta}(\theta_m,\phi_m),\nb\\
&&y =   \vec{r} \cdot \hat{y}  \simeq  D\, \hat{r}(\theta,\phi)\cdot \hat{\phi}(\theta_m,\phi_m),\nb\\
&&\dot{x} =  \vec{v} \cdot \hat{x} - \vec{v}_m \cdot \hat{x}\simeq -D\, \mu_{\delta} - \vec{v}_m \cdot \hat{x} ,\nb\\
&&\dot{y} =  \vec{v} \cdot \hat{y} - \vec{v}_m \cdot \hat{y} \simeq D\, \mu_{\alpha*} - \vec{v}_m \cdot \hat{y},
\eqn 
where we have used the orthogonality of the unit vectors to set $\vec{r}_m \cdot \hat{x}=\vec{r}_m \cdot \hat{y}=0$, $D = 147 \pm 12$~(kpc) is the distance of the Fornax dSph from us \cite{2012AJ....144....4M}, $\mu_{\delta}$ is the proper motion in declination, and $\mu_{\alpha*}$ is the proper motion in right ascension. We define $\vec{v}_m \cdot \hat{x}$ and $\vec{v}_m \cdot \hat{y}$ such that the mean of star velocities is zero, i.e., $\langle \dot{x}\rangle=0$ and $\langle \dot{y}\rangle=0$. Moreover, some of the terms have been neglected knowing that for any of the stars,  
$r\simeq D$,  
$\hat{\theta}\cdot\hat{\theta}_m \gg \hat{r}\cdot\hat{\theta}_m$,
$\hat{\theta}\cdot\hat{\theta}_m \gg \hat{\phi}\cdot\hat{\theta}_m$,  
$\hat{\phi}\cdot\hat{\phi}_m \gg \hat{r}\cdot\hat{\phi}_m$,  
$\hat{\phi}\cdot\hat{\phi}_m \gg \hat{\theta}\cdot\hat{\phi}_m$. 
Figure~\ref{Fig:UVecDotProd} shows the histograms of these dot products for the stars in the Fornax dSph, confirming the validity of the inequalities. Finally, we rotate the comoving coordinate system around its z-direction until the new x and y components of the velocities are not correlated. This rotation results in the removal of the nondiagonal components of the dispersion velocity tensor, which we can confirm explicitly when the tensor is estimated from the data. The positions and velocities of the stars in the comoving coordinate system of the Fornax dSph are shown in Figure~\ref{Fig:StarsPositionVelocity}.

Unfortunately, the dSph galaxies are so far away that the z-components of the positions and velocities of their stars cannot be inferred using Gaia. To compensate for this limitation, we assume that the phase-space density $f^*$ has the following form $f^* \rightarrow f^*(x,\dot{x},y,\dot{y})f^*_2(z,\dot{z})$. The probability distribution of stars in galaxies, i.e., $f^*$, is often approximated as the Maxwell--Boltzmann distribution; see, for example, \cite{2008gady.book.....B}, which satisfies the assumed separation. Although this assumption is widely used in the community for the mean field analysis, its validity is not known for a next-order analysis such as the one presented in this paper. This restriction, hence the assumption, can be avoided when other datasets that have the z-components of the stars are combined with the Gaia dataset, when $f^*(x,\dot{x},y,\dot{y},z,\dot{z})$ can be estimated in the full six dimensions. We leave this for future work.

\subsection{Estimation of the Phase-Space Density}
\lb{Subsec:DensityEstimation}
At this point, we would like to use the Cartesian components of the positions and velocities of the stars from the previous subsection and estimate $f^*$ of Equation~\eqref{Eq:rho_sig_vismatter}. As mentioned above, we assume that $f^*$ is only a function of the four dimensional variable $\vec{q}\equiv (x, y,\dot{x}, \dot{y})$ and the $z$ components have been integrated out. This assumption makes the following analysis a showcase of the method rather than a full analysis. The reason is that we are averaging the over- and underdensities along the $z$-axis and then measuring the correlations between the z-averages across the x--y plane. Therefore, our discovery potential will be limited to special forms of DM substructures that are not washed out by integration over the $z$ components. 
In the future, when the z-information of the stars is collected from other experiments, the same procedure leads to thorough shrinkage of the parameter space of any proposed DM scenario, and the limitation on our analysis' discovery potential will be elevated. 

We convert $\vec{q} = (x, y,\dot{x}, \dot{y})$ to dimensionless variables by dividing each of the components by the corresponding maximum value among all the stars. Next, we use the kernel density estimator \cite{10.1214/aoms/1177728190,2012arXiv1212.2812Z} to estimate $f^*$ in the four dimensional phase-space.  
The advantage of the kernel density estimator is that missing a few random member stars does not fundamentally change the estimation of the probability function but only increases its error. The method is the same as the one we used in Section~\ref{Sec:Eagle} to extract the DM mass density probability distribution with the difference that it is used to estimate the probability of the four-dimensional dimensionless phase-space 
\bqn
f^*(\vec{q}) = \sum_{i^*} \exp\left(-\frac{|\vec{q}-\vec{q}_{i^*}|^2}{h^2}\right),
\eqn
where $i^*$ enumerates the stars of the dSph.
We find that the optimal $h$ scaled back to the position-space is equivalent to $\sim $600~(pc) and scaled back to the velocities is equivalent to $\sim $38~(km/s) for Fornax dSph galaxy.

Figure~\ref{Fig:VProbability_atCenterOfFornax} shows $f^*$ at the center of the Fornax dSph. The figure indicates that the velocity distribution of stars at that position is slightly different from the Maxwell--Boltzmann distribution, which is due to the fluctuations that we aim to compute. Nevertheless, the overall Maxwellian form of the distribution confirms our assumption in Section~\ref{Sec:RealizeDM_From_f} that the Fornax dSph is in a steady state. We estimate one such distribution for every position on the x--y plane of the Fornax dSph. Having an estimation of single-body phase-space density, we insert it into Equation~\eqref{Eq:rho_sig_vismatter} to compute the density and dispersion velocity tensor at an arbitrary location $(x, y)$. To carry out the derivatives of the two objects, we repeat the estimations of the density and the dispersion velocity tensor at $(x+dx, y), (x+2\,dx,y), (x,y+dy), (x,y+2\,dy)$ and use the finite difference method. Inserting the variables and their derivatives into Equation~\eqref{Eq:DMMassDensityInTermsOfVisible} and neglecting the first term on the right-hand side gives an estimation of the mass density of DM at $(x, y)$.  
We repeat the same series of computations to estimate the DM mass densities over all the positions within a square region on the x--y plane of side length 800~(pc) centered at the origin of the Fornax dSph.

We carry out Equation~\eqref{Eq:DMMassDensityInTermsOfVisible} in a cylindrical coordinate system whose symmetry axis lies along the x-axis of the comoving coordinate system. Due to the rotation of x--y plane around the z axis, which was explained in Section~\ref{Subsec:Coordinate}, the x-axis is the symmetry axis of the system. Hence, the lack of observation of the z-component of the stars is compensated. 
The purpose of using a cylindrical over a spherical coordinate system is to avoid unnecessary additional assumptions about the x--y plane. 
More specifically, the coordinates of the cylindrical system, on the x--y plane, are $(y, \alpha, x)$, where $x$ and $y$ are the components of the comoving Cartesian system and $\alpha$ is the polar angle on the y--z plane.  
In this system, the dispersion velocity tensor takes the following diagonal form $\sigma^{2 i}_{~j}=(\sigma^{2}_{yy},  \sigma^2_{yy},  \sigma^2_{xx})$, where the indices are raised and lowered by the following diagonal metric $g_{ij}=(1,  y^2,  1)$. Since the metric is different from unity, the partial derivatives in Equation~\eqref{Eq:DMMassDensityInTermsOfVisible} shall be replaced with the covariant derivatives whose connections are the Christoffel symbols of the metric.

We estimate both the statistical and the systematic errors of our estimation. To estimate the statistical error, we use the estimated $f^*$ to draw a random sample dataset of the same size as the original one. Next, we estimate $f^*$ from the generated dataset by repeating the whole process above. We repeat the sampling and $f^*$ estimation until the standard deviation of the estimations of $f^*$ reaches steady values. The stable standard deviations are assigned as the statistical errors. The systematic errors are estimated by propagating the errors of the variables in the Gaia dataset. We use the ``uncertainties'' package in the Python programming language \cite{UncertaintiesPython} to carry out the propagation. We report that the statistical errors are negligible with respect to the systematic ones.

\subsection{Results}
\lb{Sec:Results}

So far, we have found the DM mass density $\rho$ at every position within a square on the x--y plane with a side length of $\sim 800$~(pc) whose center is at $x=y=0$.  
The estimated DM mass density distribution for the Fornax dSph can be seen in Figure~\ref{Fig:DM_massDensity_Fornax}.

We use Equations~(\ref{Eq:rho_bar}) and (\ref{Eq:varphi_Def}) to compute the fluctuations $\varphi(x,y)$ and substitute them into Equation~\eqref{Eq:FlucCorr} to estimate the correlations between fluctuations as a function of the distance $\Delta$. The two-point correlations can be seen in Figure~\ref{Fig:ObservedCorrelation}. We observe no meaningful correlations between the DM mass density fluctuations at distances $\Delta >100$~(pc).  
For $\Delta < 100$~(pc), we observe that the correlation starts to deviate from zero and increases as $\Delta$ goes toward smaller distances. This correlation is induced by the kernel density estimator rather than being genuine. Unfortunately, given the current statistics of stars, the smoothing parameters $h$ are rather large. Therefore, $f^*$ and, as a result, the DM mass density at a given point $(x, y)$ are estimated using all the stars whose distances from $(x, y)$ are smaller than $h$---the closer the stars are to the $(x,y)$ point, the more contribution to the estimation. The mass densities are expected to be correlated in distances sufficiently smaller than $h$. At distances comparable to or larger than $h$, no smoothing takes place and the estimated correlations are genuine.  
\begin{figure}
\hspace{-3mm}\includegraphics[width=.9\columnwidth]{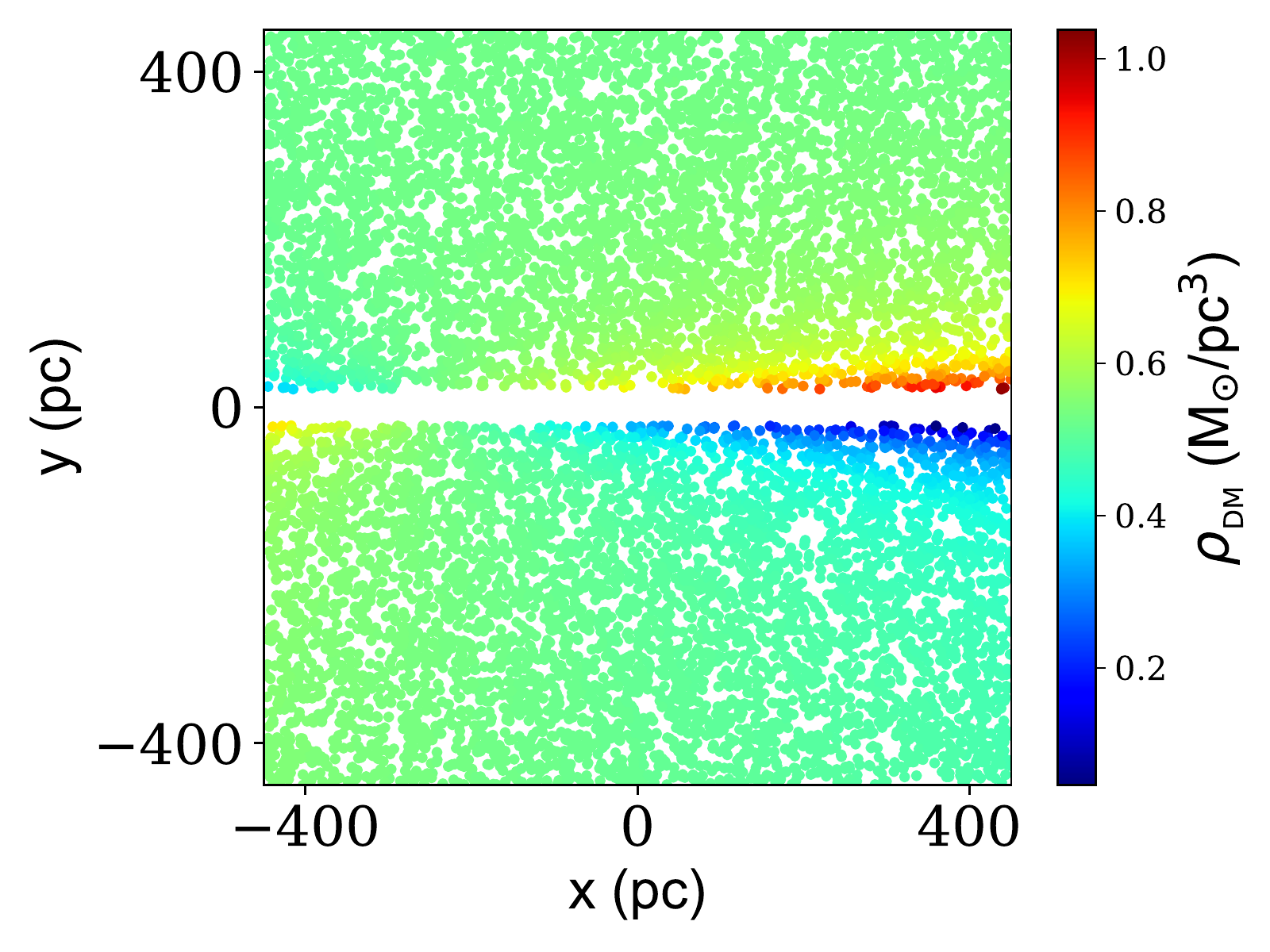}
\caption{The mass density distribution of DM at the center of the Fornax dSph galaxy on the x--y plane of the comoving coordinate system. 
We have randomly picked 10,000 points in this $x-y$ region and have calculated the DM mass density at those points. The plot indicates a core distribution of DM mass at the center of the Fornax dSph, with mean DM mass density of 0.5~$(\text{M}_{\odot}/\text{pc}^3)$. The mass density fluctuations can be readily seen. There are ingenuine correlations between the points closer than 100~(pc), i.e., similar colors grouped together, which are induced by the kernel density estimator. \lb{Fig:DM_massDensity_Fornax}
}
\end{figure}

\begin{figure}
\hspace{-2.5mm}\includegraphics[width=.9\columnwidth]{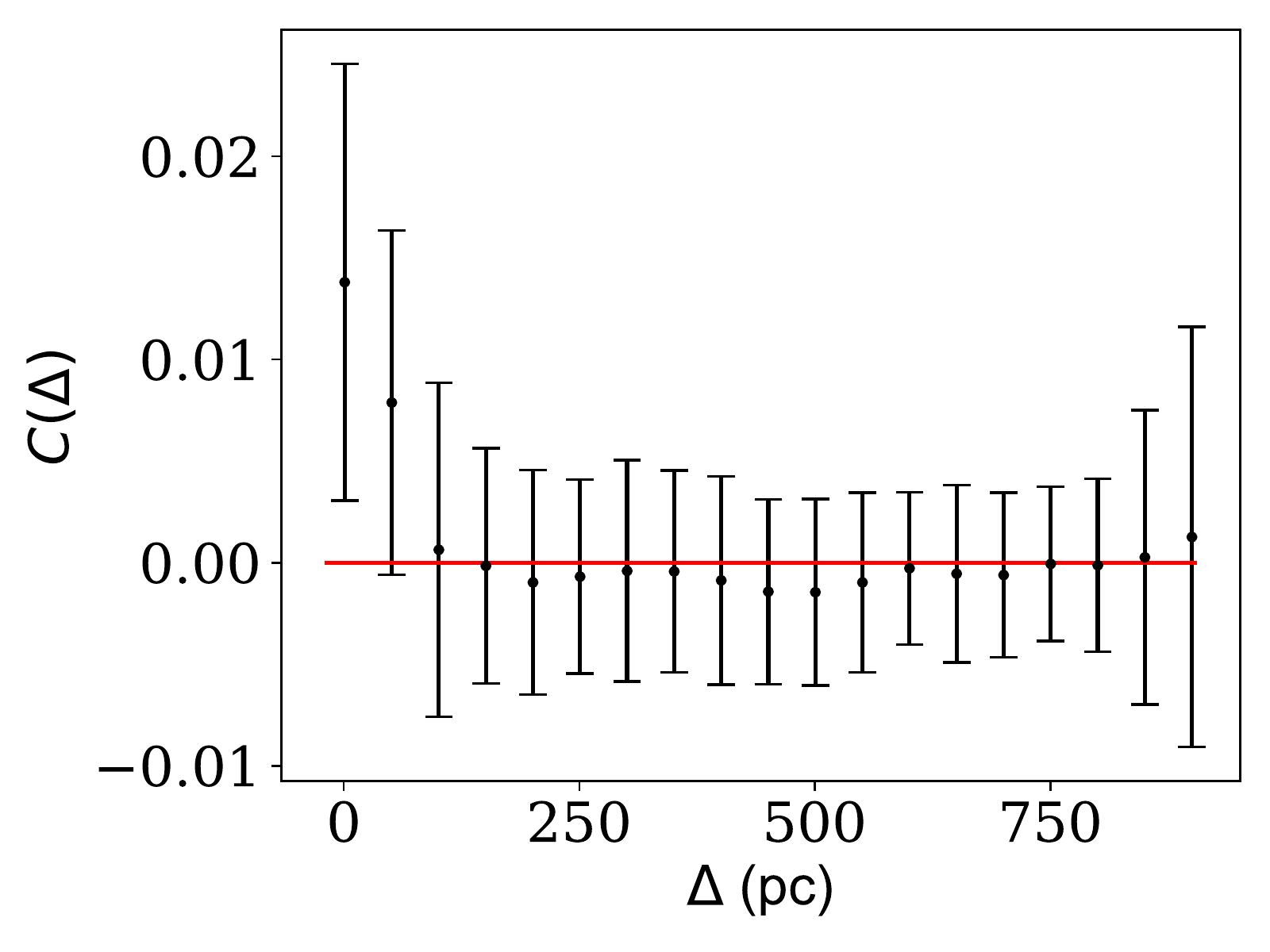}
\caption{The two-point correlations $C(\Delta)$ between DM mass density fluctuations as a function of the distance between the position of mass densities $\Delta$ in the Fornax dSph. Correlations at $\Delta < 100$~(pc) are induced by the kernel density estimator and are not genuine.  
We observe no statistically significant correlation at larger distances in the galaxy.   \lb{Fig:ObservedCorrelation}}
\end{figure}

At this point, we can use the results of Figure~\ref{Fig:ObservedCorrelation} to restrict the $\beta\gamma-\beta\nu^2$ parameter space of the free energy of Section~\ref{Subsec:TheoryDensityCorrs}. We use the left-tailed $\chi^2$ method at a 5\% significance level  to place the limitations. The solid shaded region in Figure~\ref{Fig:RuledOutRegions} is excluded for those models with $\nu^2 > \gamma$. Both the solid and the hatched shaded regions are excluded for models with $\nu^2 \ll \gamma$. The white region to the right of the Figure awaits future investigations. In the following, we discuss two DM models that belong to the two categories.   

\begin{figure}
\hspace{-2.5mm}\includegraphics[width=\columnwidth]{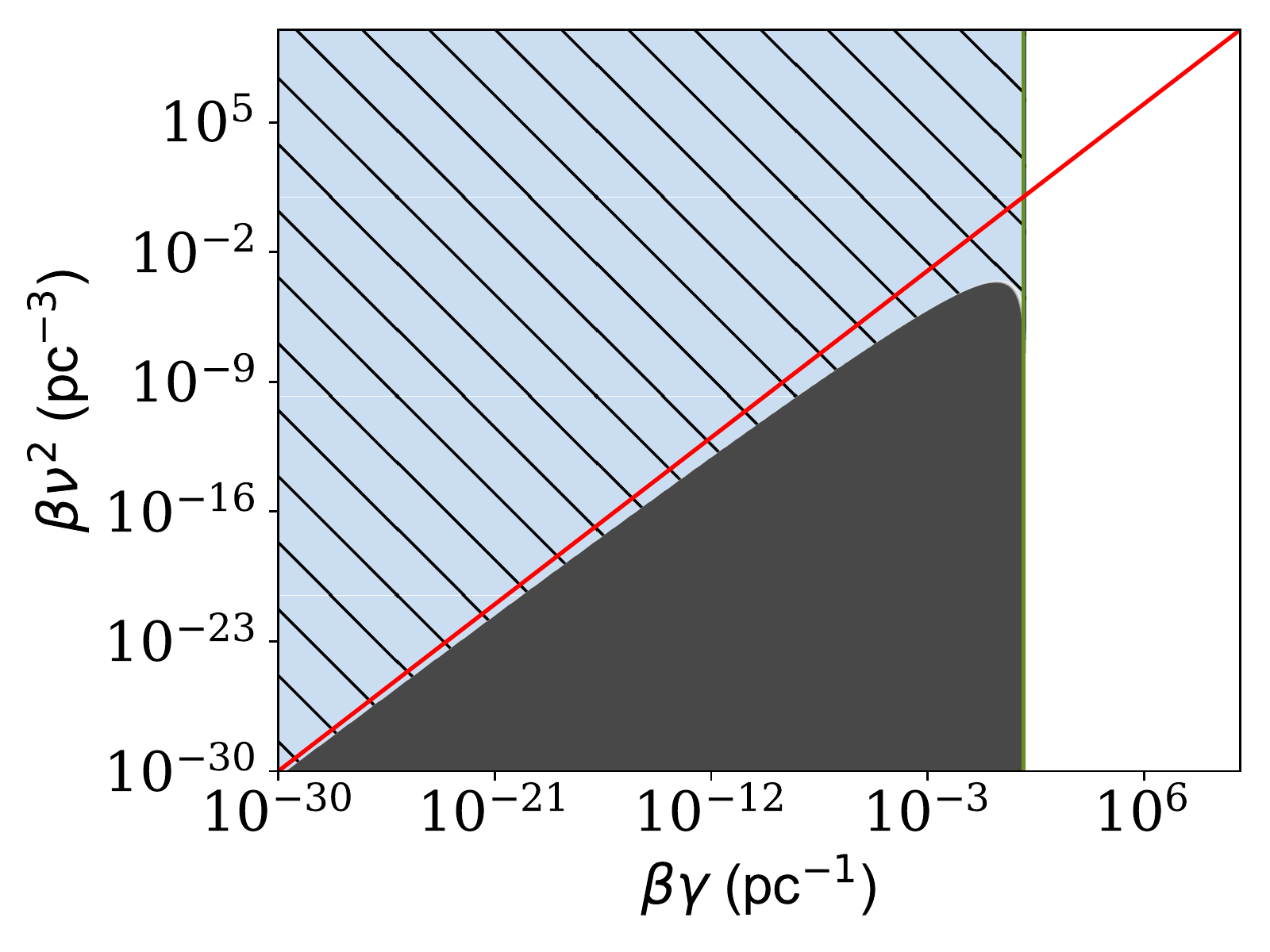}
\caption{Bounds on the coefficients of the free energy expansion at 5\% significance level using $\Delta > 100~$(pc) of Figure~\ref{Fig:ObservedCorrelation}. The vertical line is at $\beta \gamma = 10 ~(\text{pc}^{-1})$ and the diagonal red line shows $\beta\gamma=\beta\nu^2$. 
The solid gray region is excluded for DM models that live far from their critical temperature. An example is a dark classic gas. The solid grey region and the hatched blue region, i.e., the entire region left to the vertical line, are excluded for DM models that live close to their critical temperature. Superfluid models of DM for example.   
In any model of DM, there is a map between $\gamma$ and $\nu^2$ parameters and the underlying principles of the model. Therefore, this plot can be used to explore the allowed regions of the parameter space of DM models.  \lb{Fig:RuledOutRegions}}
\end{figure}

\paragraph*{A Classic Weakly Collisional Gas of DM}
We assume a gas of DM with weak gravitational collisions between its particles. The interaction between two dark particles is assumed to be a function of their distance $r$ and its potential energy is $u(r)$. Defining $u_0\equiv\int d^3x\, u(r)$ and $u_2\equiv \int d^3x\,r^2u(r)$, the coefficients of the free energy expansion in Equation~\eqref{Eq:GeneralPartionFunc} are 
\bqn 
\lb{Eq:SampleDMModelCoefs}
&&\beta\gamma = \beta\frac{\bar{\rho}^2}{m^2} u_2, \nb\\
&&\beta\nu^2 = \frac{\bar{\rho}}{m}\left(1 + 2\beta  \frac{\bar{\rho}}{m} u_0\right),
\eqn
where $m$ is the mass of DM particles and the collisions are assumed weak enough that $\gamma < \nu^2$. See Appendix~\ref{App:InteractiveGas} for the derivation.  

To place a bound on $u(r)$ using the limitations of Figure~\ref{Fig:RuledOutRegions}, we note that at the center of small galaxies, the DM temperature to mass ratio is approximately equal to $10^{-6}~$(K/eV)~\cite{2021JCAP...08..023B}. On the other hand, from Figure~\ref{Fig:DM_massDensity_Fornax}, the average mass density of DM at the center of the Fornax dSph is $\sim$$0.5~(M_{\odot}/\text{pc}^3)$. Therefore, if DM mass is around 1~MeV, $\bar{\rho}/m$$\sim$$10^{59}(\text{pc}^{-3})$. Substituting this number into Equation~\eqref{Eq:SampleDMModelCoefs}, we can conclude that for a positive interaction, $\beta \nu^2$ is well above the excluded region in Figure~\ref{Fig:RuledOutRegions}.
Therefore, a classical gas model of DM with no or positive $u_0$ can explain the nonexisting correlations of Figure~\ref{Fig:ObservedCorrelation}. It should be emphasized that this simplistic model of DM by no means represents the sophisticated CDM model and the collisions due to baryonic feedback. However, the CDM model is still based on the evolution of a cold and classic gas. Therefore, the agreement between the analytic result presented here and the correlations predicted by the Eagle simulation of CDM presented in Figure~\ref{Fig:Correlation_Eagle} is not a surprise.

\paragraph*{DM Models at the Critical Temperatures}
Let us now focus on a scenario in which DM in small halos is in the superfluid state where the DM temperature is close to its critical point $T$$ \sim$$ T_c$. It is known that close to the critical temperature, the correlation length diverges in the following form $\gamma/\nu^2$$ \sim$$ |T-T_c|^{-1}$. As a result, the exponential in Equation~\eqref{Eq:Theoretical_Correlation} disappears and the correlation function takes the form of 
\bqn
\lb{Eq:Superfluid_Correlation}
C(\Delta) = 4\pi (\beta\gamma\Delta)^{-1}.
\eqn
One can observe that $\beta \nu^2$, which was somewhat related to the number density of particles in Equation~\eqref{Eq:SampleDMModelCoefs}, is absent in this equation. Using the $\chi^2$ test, we can conclude that only $\beta\gamma > 10~(\text{pc}^{-1})$ is allowed in Figure~\ref{Fig:RuledOutRegions}.

\section{Conclusions}
\lb{Sec:Conclusion}
In this paper, we have used the simulations by the Eagle project and observations made by Gaia to estimate the mass density distribution of DM within the central part of (i) a simulated small halo and (ii) the Fornax dSph galaxy. For the two mentioned halos, we have computed DM mass density fluctuations as well as the correlations between them in the same regions. We have shown that the correlations between DM mass density fluctuations are not significantly different from zero when they are $>$$ 100~$(pc) apart in any of the two halos.

Our estimation of DM mass density correlations imposes restrictions on any proposed model of DM. Moreover, since correlations between density fluctuations are independent of the mean density, these limitations are in addition to those applied by observations of mass profiles of DM, through rotation curves, for example. We foresee two approaches for imposing restrictions on DM models using the estimation of mass density correlations from observations. In the first approach, provided that high-resolution N-body simulations exist, one can use Equation~\eqref{Eq:FlucCorr} to compute the predicted correlations using simulations and compare them with Figure~\ref{Fig:ObservedCorrelation}. This route has been taken in this paper in Section~\ref{Sec:Eagle}. 

In the second approach, one writes down the general form of the statistical field theory of DM mass density. Assuming that the DM halo can exchange dark particles with its host and/or cosmic background, halo's stability can be assumed, and gravothermal catastrophe is prevented by the induced chemical potential. Therefore, the free energy of the halo can be expanded around its steady state and the perturbation theory can be used to calculate the correlations between DM mass densities without knowing the underlying physics of the DM scenario. We have used Figure~\ref{Fig:ObservedCorrelation} to lay bounds on the coefficients of the free energy expansion. Since the coefficients are functions of the physics of the given DM model, the limitations can then be propagated into the model's parameter space. We have used this approach to explore the parameter space of (i) a gas model of DM with weak collisions between its particles, and (ii) a superfluid DM at its critical temperature. The excluded regions of their parameter spaces have been presented.

The data analysis of this paper can be improved in several aspects. The radial velocities and distances of the member stars of dSph galaxies have been measured in other experiments. A combination of their datasets with Gaia observations can help avoid unnecessary assumptions. Such combinations might also help with identifying more member stars, which would subsequently increase the statistics and improve, or decrease, the smoothing parameter $h$. The density of observed stars is not uniform across the halos. Especially, there are many more stars at the center of galaxies than at large distances from the centers. Hence, the analysis can enjoy an adaptive $h$ estimation with better resolution in regions where more stars have been observed. Since no public well-established  implementation of this adaptive smoothing is available, we have left it for future work. Finally, new experiments targeting higher statistics and better precision can help to explore smaller length scales and tighten the limitations on DM models.






\appendix

\section[Classic Gas Model of DM]{Classic Gas Model of DM}

\lb{App:InteractiveGas}
The coefficients of the expansion of the free energy functional in Equation~\eqref{Eq:GeneralPartitionFunction} are related to the underlying physics and serve as the fingerprints of a proposed DM scenario. In this section, we would like to use a simple example to demonstrate the map between the DM model and these coefficients.

We start by writing the total energy of a classic DM gas in the halo 
\bqn
E = \sum_i E_i + \sum_{ij} u_{ij},
\eqn
where $i$ and $j$ enumerate dark particles, and $u_{ij}$ is the effective potential energy of the gravitational collisions between the particles. Since $\rho(x) = \sum_i m \delta^3(\vec{x}-\vec{x}_i)$, where $m$ is the mass of dark particles, the sum over the particles can be replaced by $\sum_i = \int d^3x\, \frac{\rho(x)}{m}$. Therefore, in a grand canonical ensemble, the free energy functional reads
\bqn
\lb{Eq:Free_Energy_InteractiveGas}
\beta F[\rho] &=&
\beta F_{\text{free}}+
\int d^3x
 \left( 
 -\beta \frac{\rho}{m}\mu_f + J\rho\nb
 \right)\nb\\
 &&+
 \beta \int d^3x\, d^3x' \frac{\rho(x)}{m}\frac{\rho(x')}{m} u(|\vec{x}-\vec{x}'|),
\eqn
where $\mu_f \equiv \left(\mu - m\phi \right)$, and the terms related to an ideal gas read \cite{LUTSKO2006229}
\bqn
F_{\text{free}}= \int d^3x \bigg( \frac{\rho}{m}\text{Ln}\left(\frac{\rho}{m}\lambda^3\right) - \frac{\rho}{m}\bigg),
\eqn
with $\lambda = \frac{h}{\sqrt{2\pi m k T}}$ where $h$ is the Planck constant. See Appendix~\ref{App:CollisionLess_FreeEnergy} for the detailed derivation of $F_{\text{free}}$ and the second integral on the right-hand side of Equation~\eqref{Eq:Free_Energy_InteractiveGas}.  

At this point, we replace $\rho \rightarrow \bar{\rho}(1+\varphi)$ and separate the terms according to the powers of $\varphi$. The zeroth-order terms appear as a normalization factor and will be canceled out later. The first and second order terms read
\bqn
&&\beta F_1 = \int d^3x\, \varphi \, \frac{\bar{\rho}}{m} \bigg( \text{Ln}\bar{\rho} + \text{Ln}\frac{\lambda^3}{m} -\beta\mu_f 
 + 2\beta \frac{\bar{\rho}}{m}  u_0\bigg),
\nb\\
&& \beta F_2 = \int d^3x\, \bigg(
\frac{1}{2}\frac{\bar{\rho}}{m} \varphi^2 +\beta \frac{\bar{\rho}^2}{m^2}\varphi \int d^3x'\, \varphi(x') u(|\vec{x}-\vec{x}'|)
\bigg),\nb\\
\eqn
where $u_0 \equiv \int d^3x' \,   u(|\vec{x}-\vec{x}'|)$. 

Our assumption regarding the steadiness of the halo implies that $\frac{\delta F[\varphi]}{\delta \varphi}|_{\varphi=0}=0$, which means that the terms inside the integral in $F_1$ is zero. Hence, the steady condition requires that $\mu_f$ takes the following form
\bqn
\lb{Eq:mu_f_classicGas}
 \mu - m\phi = 2\frac{\bar{\rho}}{m}u_0 + \frac{1}{\beta}\text{Ln}\frac{\bar{\rho}\lambda^3}{m}.
\eqn
It should be noted that $\mu_f$ has to be negative for the free energy to have a minimum value corresponding to a stable steady state. If the number of particles in the system was conserved, the chemical potential $\mu$ on the left-hand side was absent and $\mu_f$ would be positive due to the negativeness of the gravitational potential. The latter is the origin of the gravothermal catastrophe in systems with dominant gravitational effects whose number of particles is not maintained by an external bath. The presence of the chemical potential on the left-hand side of Equation~\eqref{Eq:mu_f_classicGas} allows us to assume that DM halos of dSph galaxies have negative chemical potentials that establish their stable steady states.

Assuming that $\varphi$ does not vary aggressively over the effective range of $u(|\vec{x}-\vec{x}'|)$, we use the so-called square gradient approximation, do an expansion up to the second order, and insert it into the integral in the second term of $F_2$
\bqn
\int d^3x'\, \varphi(x') u(|\vec{x}-\vec{x}'|) = \varphi(x) u_0 + \frac{1}{2}\delta^{ij} \partial^2\varphi u_2, \nb\\
\eqn
where $u_2\equiv \int d^3x' \,  (|\vec{x}-\vec{x}'|)^2 u(|\vec{x}-\vec{x}'|)$.   
Therefore, the effective free energy reads
\bqn
\beta F[\varphi] = \frac{\beta}{2}\int d^3x\, \bigg(\gamma \varphi \partial^2 \varphi + \nu^2 \varphi^2 \bigg),
\eqn
where the $\gamma$ and $\nu^2$ are 
\bqn 
&&\gamma = \frac{\bar{\rho}^2}{m^2} u_2, \nb\\
&&\nu^2 = \frac{\bar{\rho}}{m\beta}\left(1 + 2\beta  \frac{\bar{\rho}}{m} u_0\right).
\eqn

\section[Free energy functional of an ideal gas]{Free Energy Functional of an Ideal Gas}
\renewcommand{\theequation}{B\arabic{equation}} \setcounter{equation}{0}
\lb{App:CollisionLess_FreeEnergy}
We begin with a sufficiently small volume interval $\delta V$ at position $\vec{x}$ of the halo with total energy $E(x)$, the chemical potential $\mu(x)$, that contains $N(x) = \frac{\rho(x)}{m} \delta V$ number of particles.  
We assume that this system is in the steady state such that its state's probability is 

\bqn
{\cal{P}}_x = e^{-\beta \left(E(x) - \mu(x) N(x)\right)}.  
\eqn

If the number of particles in the energy state $\varepsilon$ is $n_{\varepsilon}$, the total energy of the system and total number of particles read
\bqn
&&E(x) = \sum_{\varepsilon}n_{\varepsilon} \varepsilon + N(x)m\phi(x),\nb\\
&&N(x) = \sum_{\varepsilon}n_{\varepsilon} .
\eqn

For the case of an ideal gas, no correlation does exist between different locations, making it a trivial case.  
Hence, the probability of finding the halo in a particular state reads
\bqn
{\cal{P}} = {\cal{P}}_{1} {\cal{P}}_{2}{\cal{P}}_{3}\cdots,
\eqn
where ${\cal{P}}_{i} = {\cal{P}}_{x_i}$.
The partition function of the halo is the sum over all the probabilities
\bqn
{\cal{Z}}&=&\left(\sum_{N_1,E_1}\sum_{N_2,E_2} \cdots\right){\cal{P}}_1 {\cal{P}}_2\cdots
= \prod_{x} \sum_{N_x}\sum_{E_x} {\cal{P}}_{x}\nb\\
&=& \prod_{x} \sum_{N_x} e^{\beta N(x)\left(\mu(x)-m\phi(x)\right)}\frac{1}{N(x)!}\nb\\
&&\left(\frac{2\pi\delta V}{h^3}(2m)^{\frac{3}{2}}\int_0^{\infty} d\varepsilon\,\varepsilon^{\frac{1}{2}} e^{-\beta \varepsilon}\right)^{N(x)}\nb\\
&=& \prod_{x} \sum_{N_x} e^{\beta N(x)\left(\mu(x)-m\phi(x)\right)}\frac{1}{N(x)!}
\left(\frac{\delta V}{\lambda^3}\right)^{N(x)}. 
\eqn
We note that the last two terms can be re-expressed in the following form
\bqn
\frac{1}{N!}\left(\frac{\delta V}{\lambda^3}\right)^{N} = 
\exp\left(-N\,\text{Ln}\left(\frac{\lambda^3\rho}{m}\right)+N\right), 
\eqn
where we have used $a=e^{\text{Ln}(a)}$, the Stirling's approximation for Ln$(N!)$, and $\frac{N}{\delta V}=\frac{\rho}{m}$.
Hence, the partition function of the halo reads
\bqn
{\cal{Z}} &=& \prod_x \int d\rho(x)\,\exp\left(-\delta V \frac{\rho}{m}\left(\text{Ln}\frac{\lambda^3\rho}{m}-1-\beta\mu_f\right)\right)\nb\\
&=&\int D\rho \exp\left(-\int d^3x\, \frac{\rho}{m}\left(\text{Ln}\frac{\lambda^3\rho}{m}-1-\beta\mu_f\right)\right) ,
\eqn
where we have set $\sum_x \delta V \rightarrow \int d^3x$.

\section[Additional Figures]{Additional Figures}

\lb{App:AdditionalFigures}
This section presents a few additional figures.

\begin{figure}[t]
\hspace{-1mm}\includegraphics[width=0.47\columnwidth]{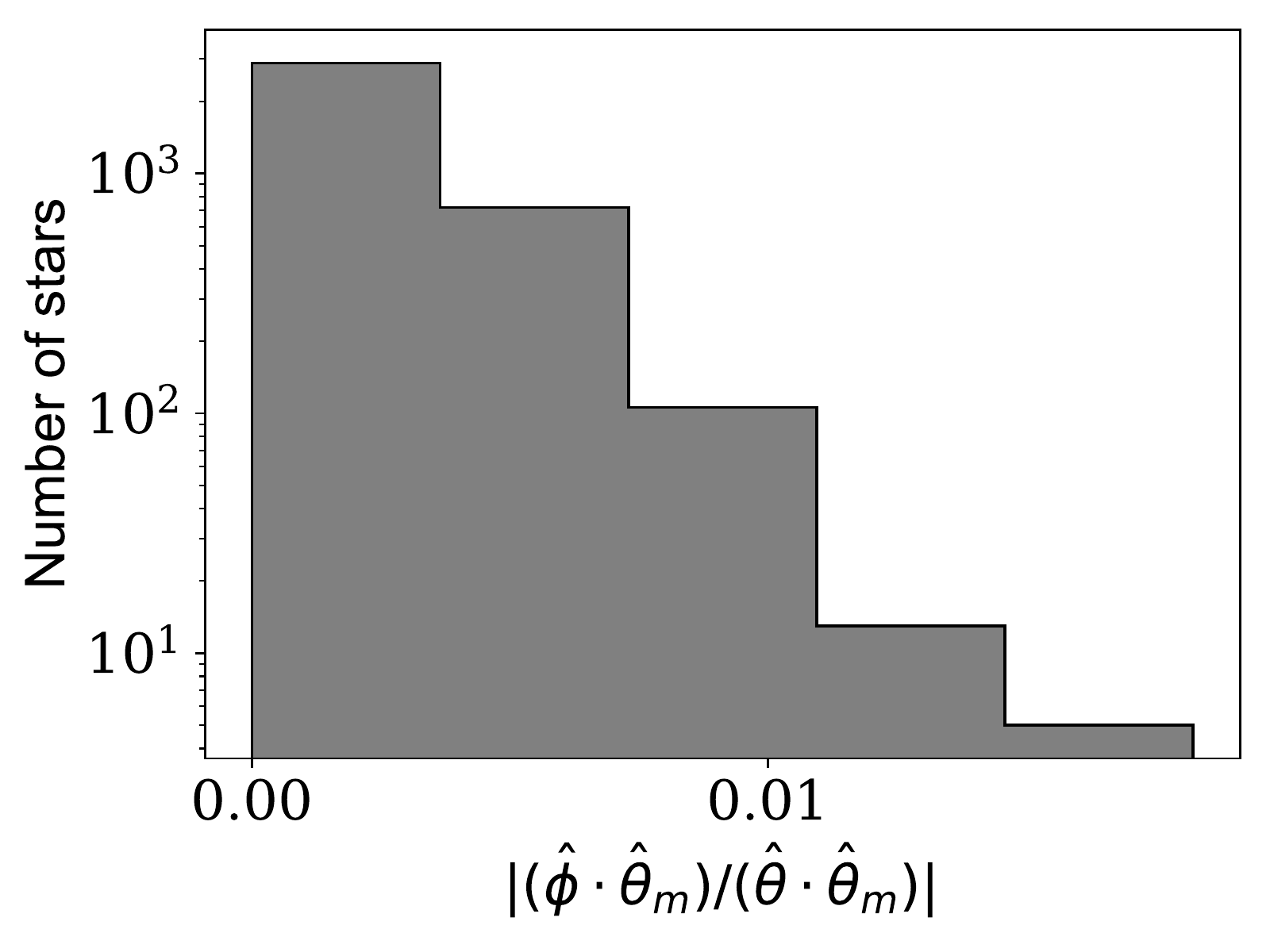}
\includegraphics[width=0.47\columnwidth]{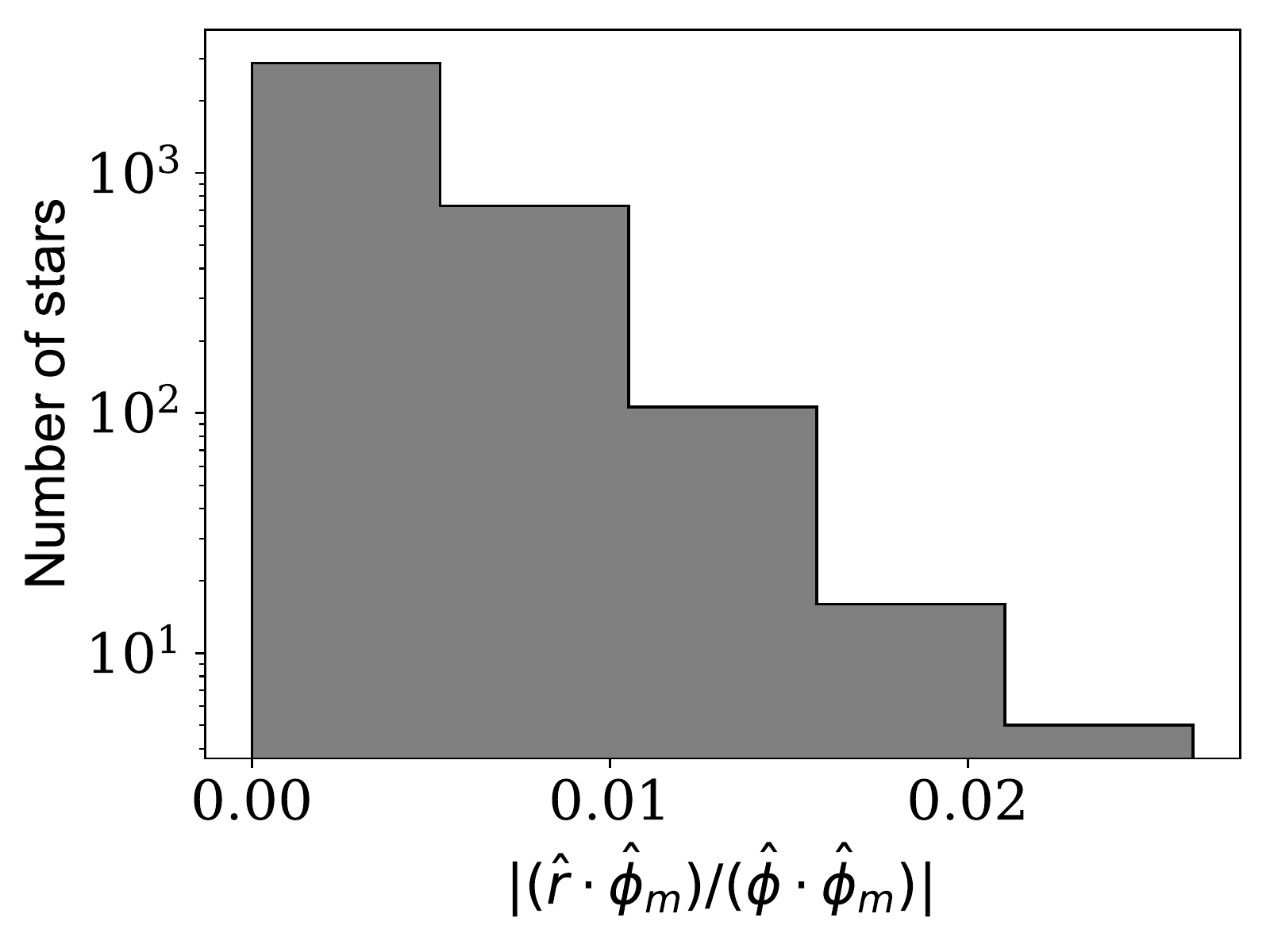}  
\caption{\textit{Cont}.}
\end{figure}

\begin{figure}[t]
\hspace{-1mm}\includegraphics[width=0.47\columnwidth]{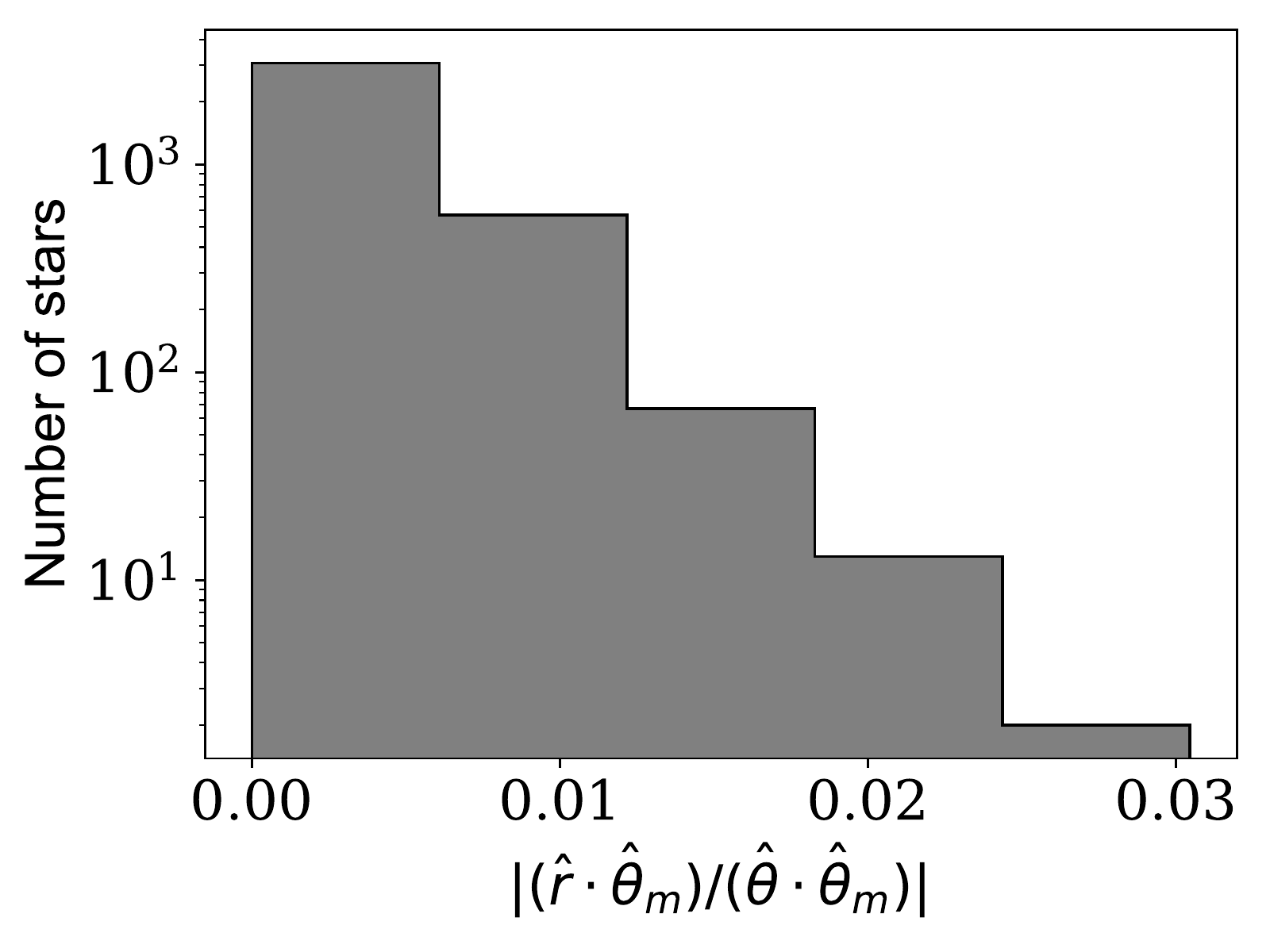}     
\includegraphics[width=0.47\columnwidth]{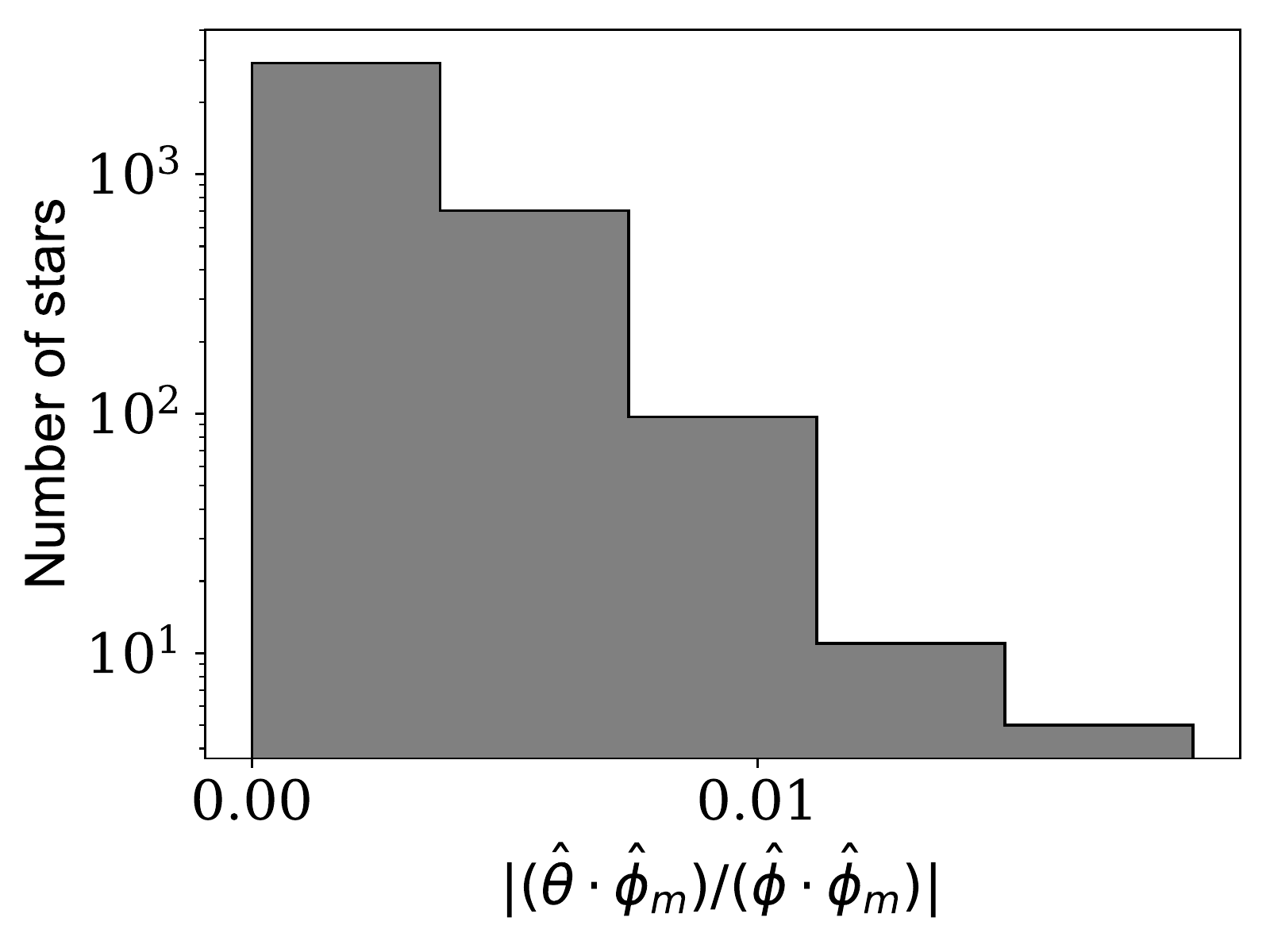}

\caption{The ratio of the dot products of the unit vectors in the spherical coordinate system for the stars that belong to the Fornax dSph as reported by Gaia. The figure indicates that all of the ratios are negligible for any of the stars.    \lb{Fig:UVecDotProd}}
\end{figure}

\begin{figure}[t]
\centering
\includegraphics[width=0.49\columnwidth]{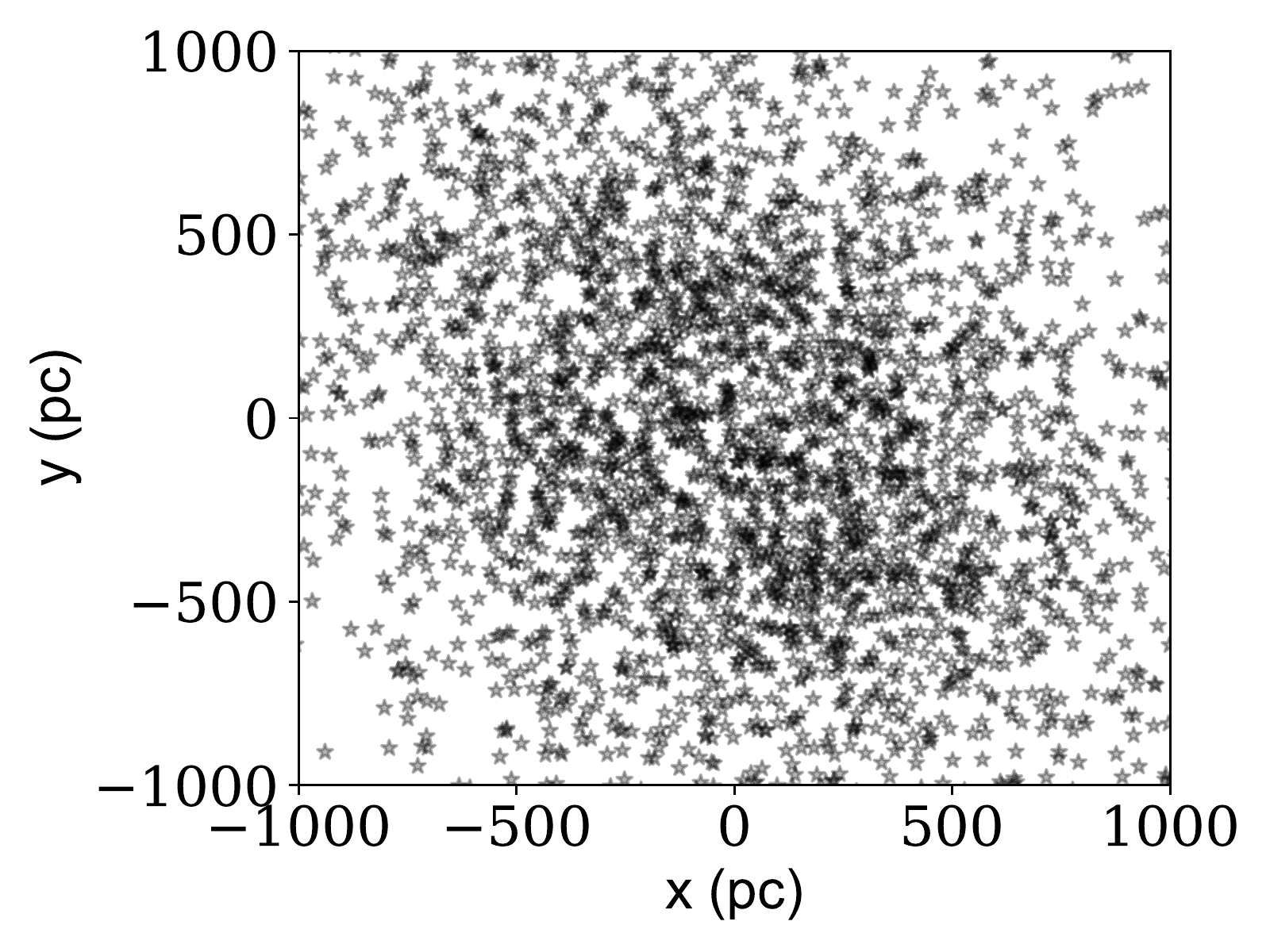}
\hspace{-1mm}\includegraphics[width=0.49\columnwidth]{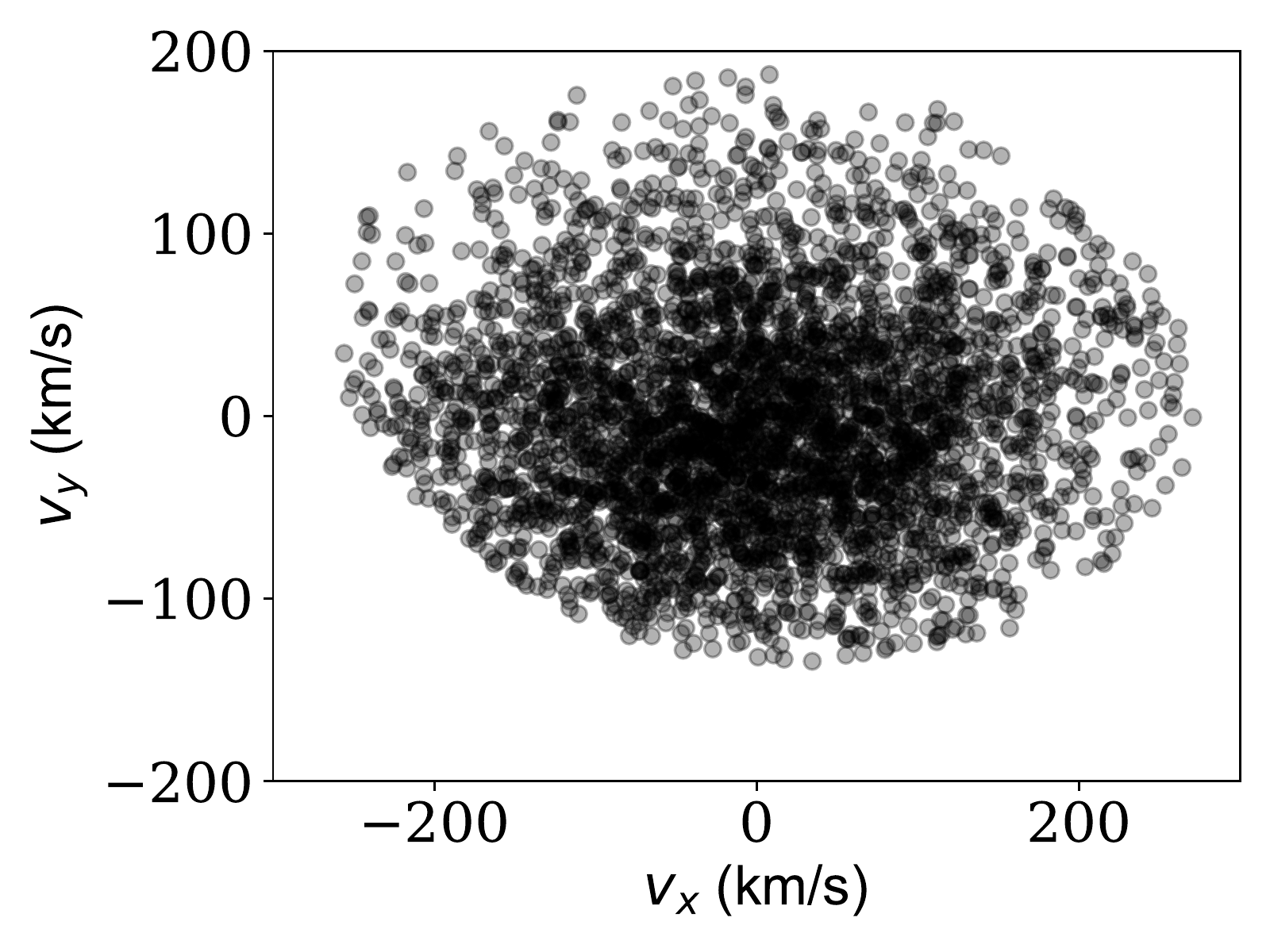}  
\includegraphics[width=0.49\columnwidth]{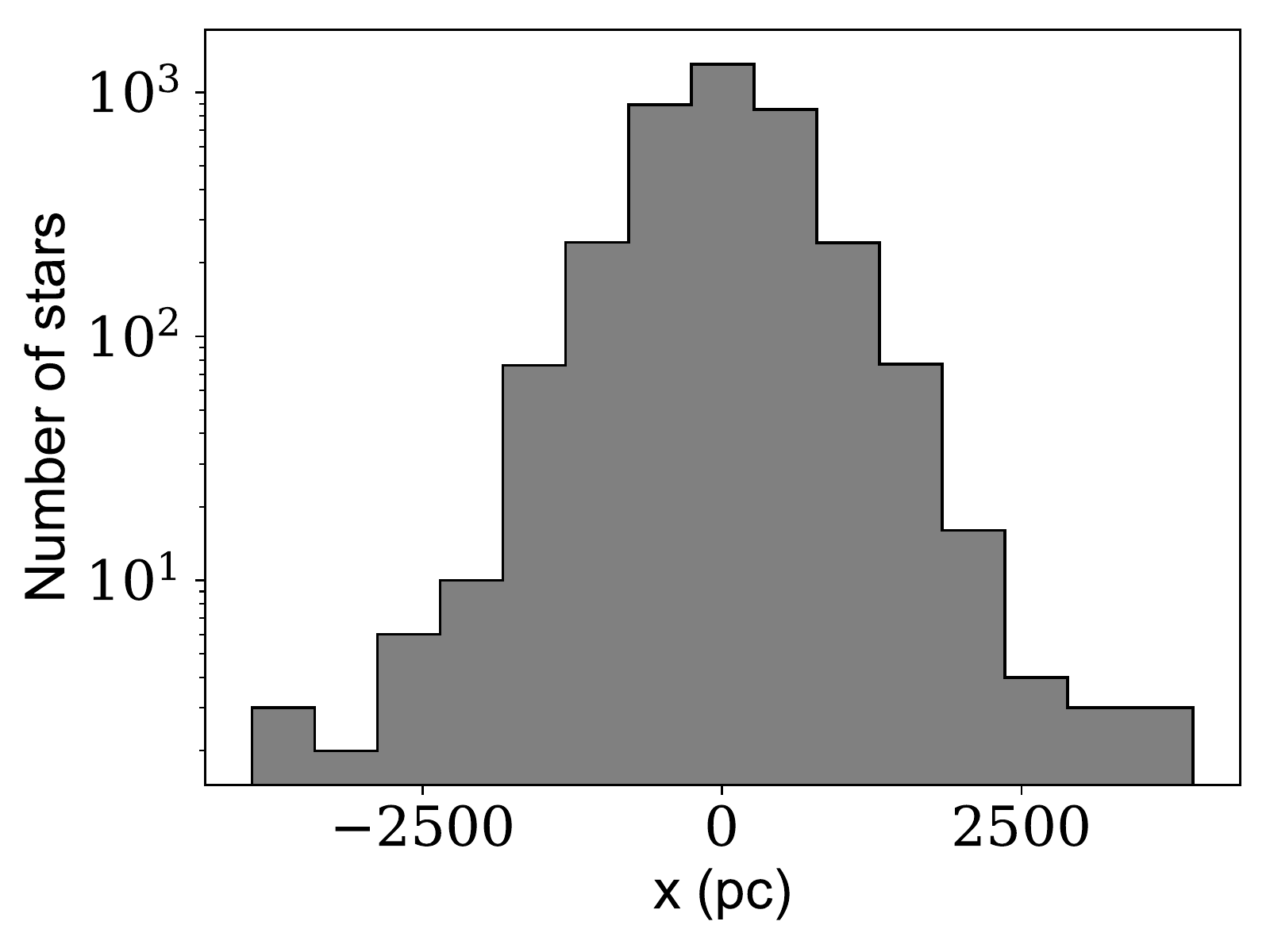} 
\includegraphics[width=0.49\columnwidth]{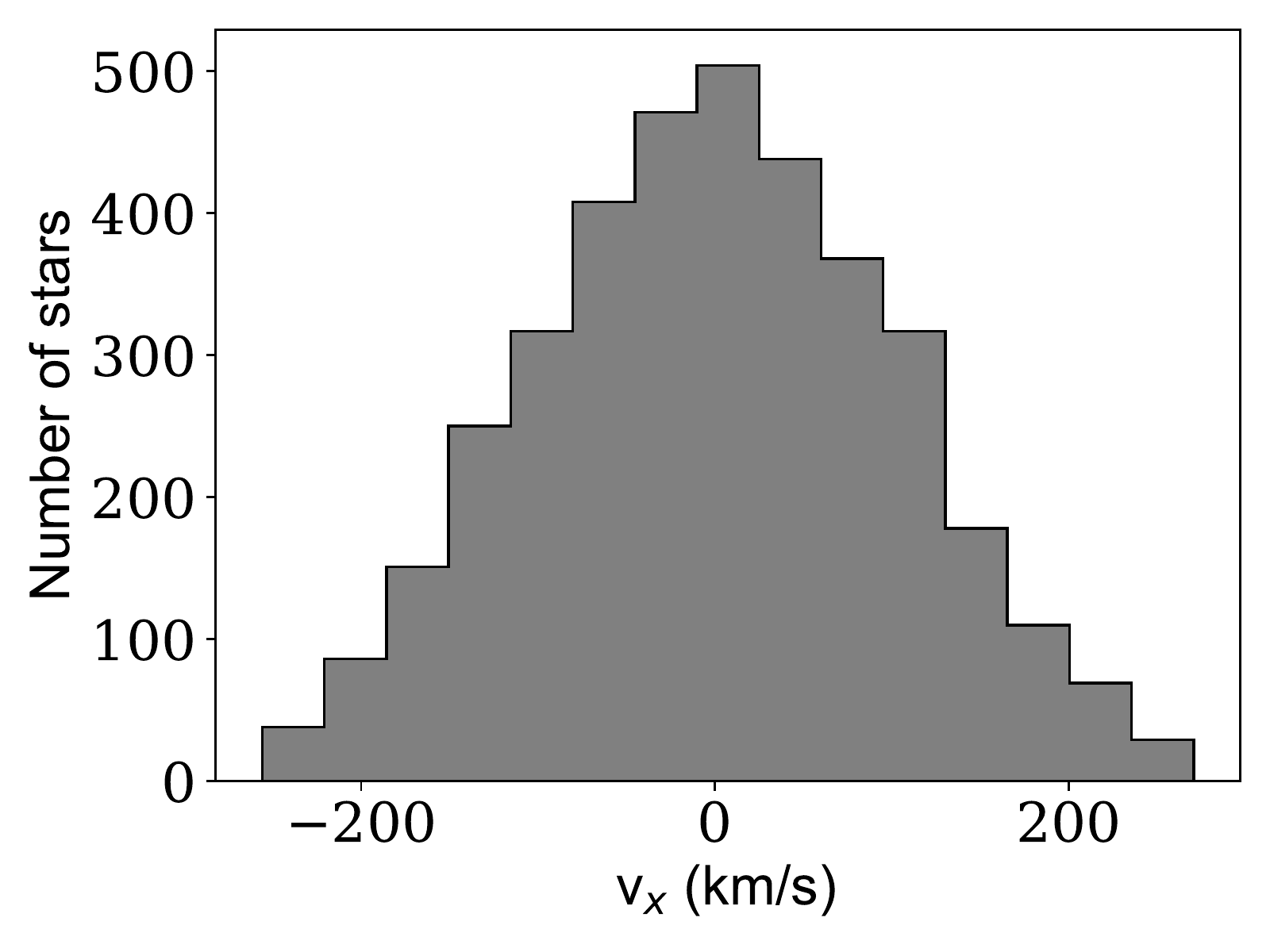}
\caption{\textit{Cont}.}
\end{figure}

\begin{figure}[t]
\includegraphics[width=0.49\columnwidth]{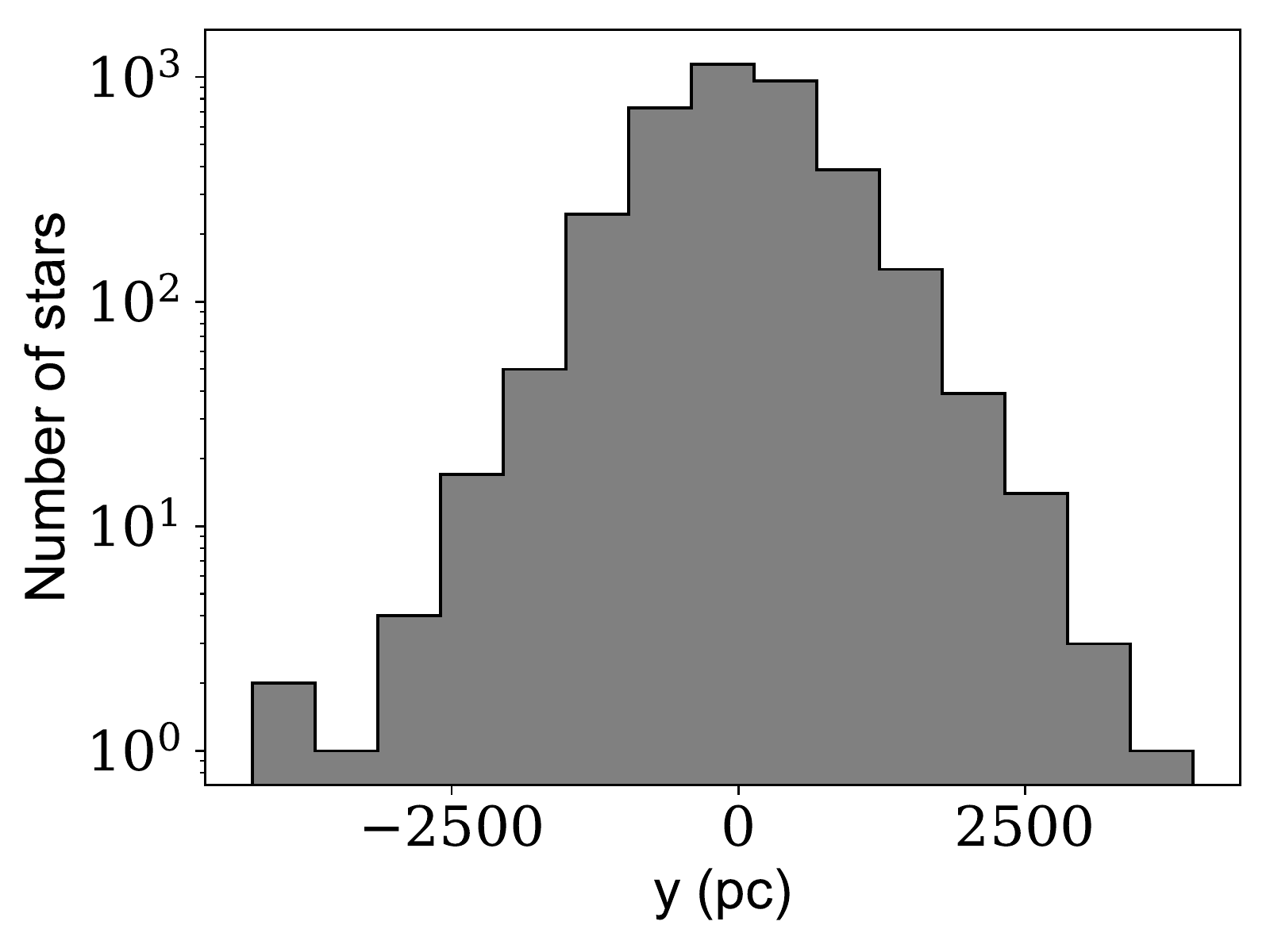}
\includegraphics[width=0.49\columnwidth]{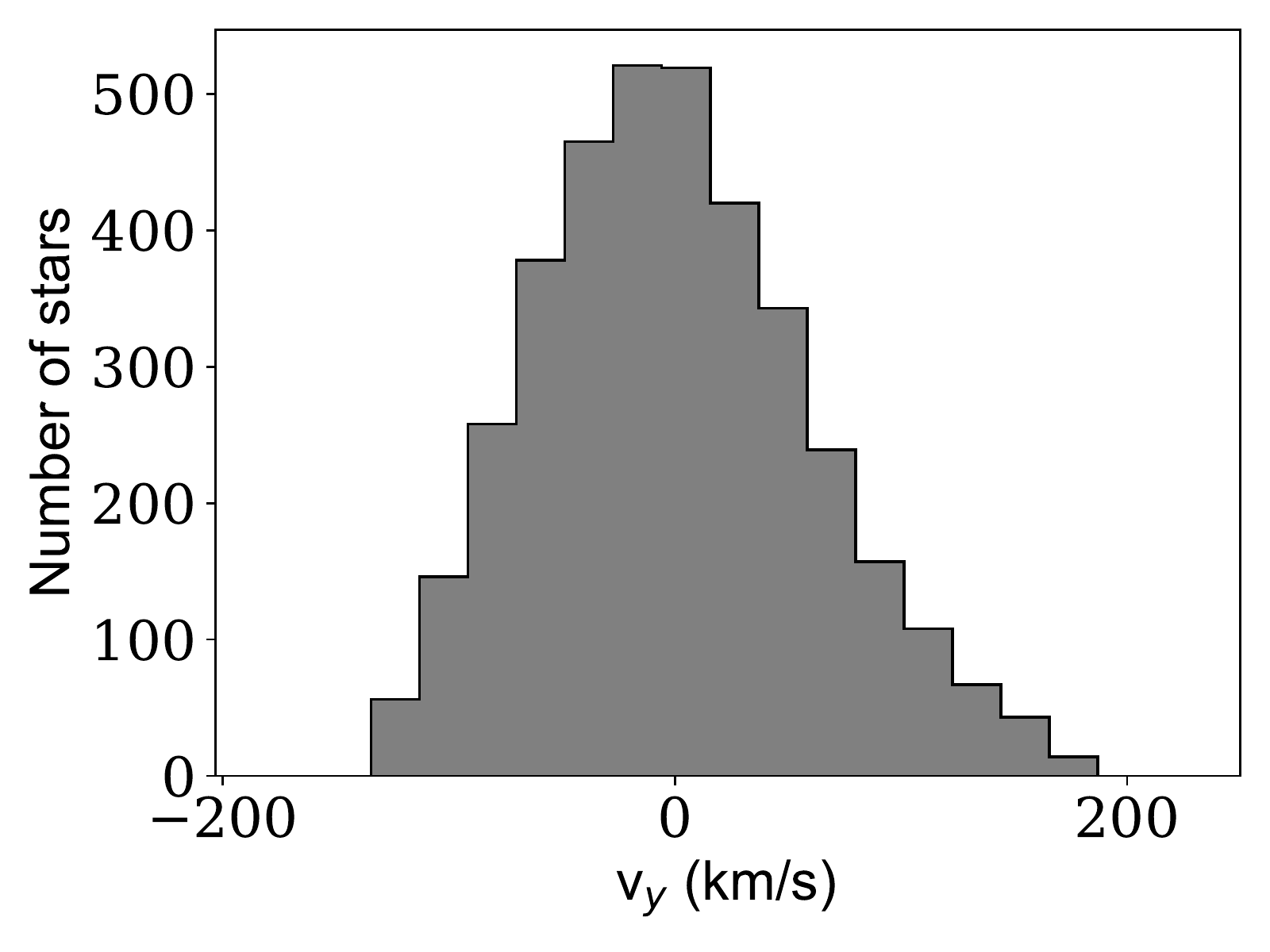} 
\caption{The position and velocity of the stars in the comoving coordinate system of the Fornax dSph. Each point in the two top plots represents a start. The rest of the plots are the histograms of the four dimensions of the phase-space.
The histograms indicate that the system is fairly spherical. The height and the width of the histograms of $x$ and $y$ are approximately equal supporting the spherical symmetry in the position-space. The same is approximately valid for $v_x$ and $v_y$ histograms supporting the spherical symmetry in the velocity-space. The reason we have used a cylindrical symmetry to carry out the Jeans equations in Section~\ref{Sec:Data} is the slight discrepancy between the width, but not the height, of $v_x$ and $v_y$ histograms.   \lb{Fig:StarsPositionVelocity}}
\end{figure}

\begin{figure}[t]
\hspace{-4mm}\includegraphics[width=\columnwidth]{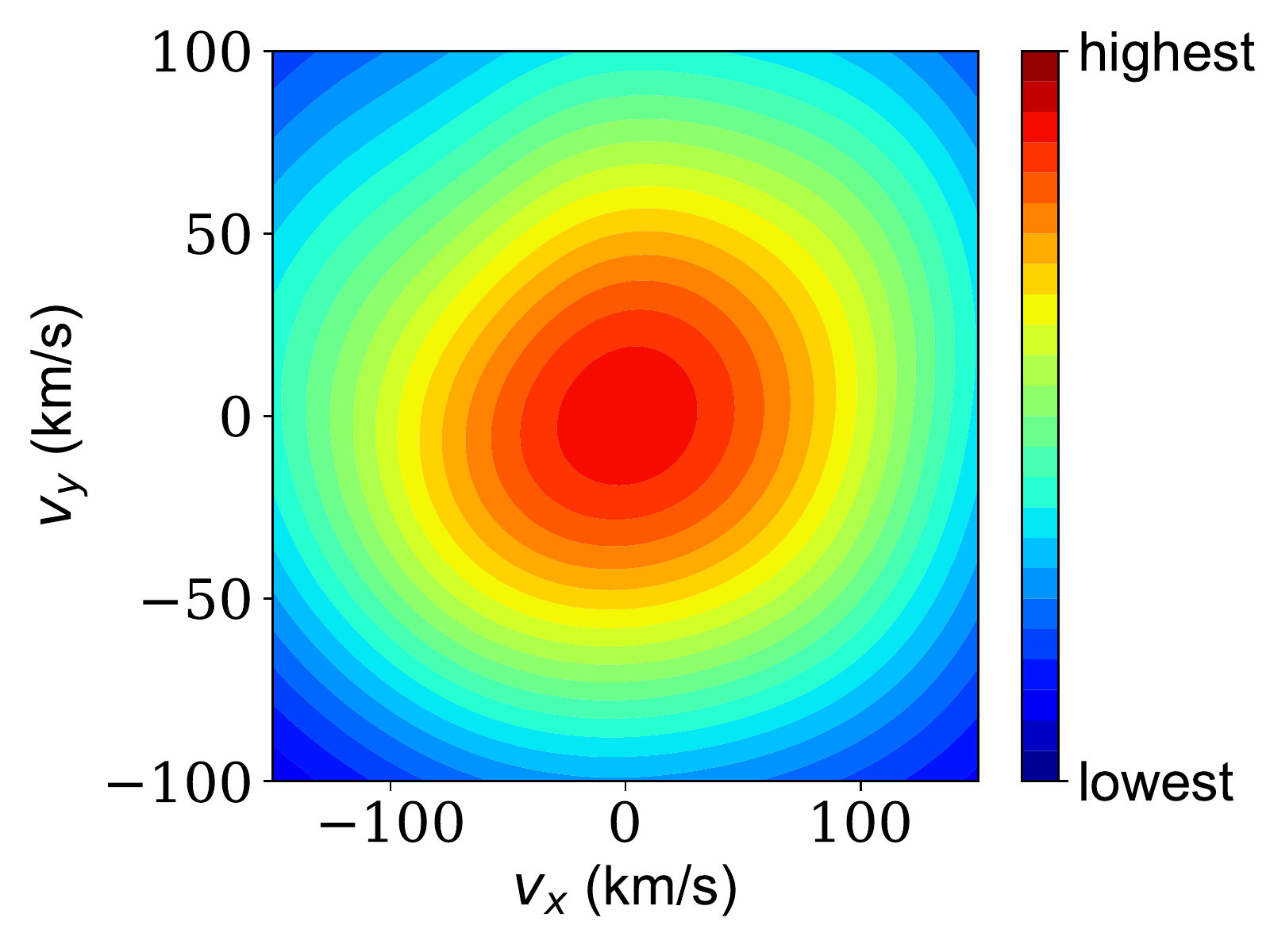}
\caption{The probability distribution of stars' velocities at the center, $x=y=0$ in the co-moving coordinate system, of the Fornax dSph estimated using the kernel density estimator as described in Section~\ref{Sec:Data}. The distribution is close to the Maxwell--Boltzmann distribution function although a slight deviation can be seen. The deviations are slightly different at other $(x, y)$ positions leading to the mass density fluctuations of Figure~\ref{Fig:DM_massDensity_Fornax}.\lb{Fig:VProbability_atCenterOfFornax}}
\end{figure}

\newpage

\end{document}